\documentclass{elsarticle}
\usepackage{hyperref}

\journal{Journal of \LaTeX\ Templates}

\usepackage{amssymb}
\usepackage{graphicx}
\usepackage{url}
\usepackage{amsmath}
\usepackage{amssymb}
\usepackage{color}
\usepackage{epsf}
\usepackage{color}
\usepackage{graphicx}
\usepackage{epsfig}
\usepackage{fancybox}
\usepackage{multicol}
\usepackage{hhline}
\usepackage{xspace,epic,eepic,graphicx}
\usepackage{latexsym}
\usepackage[all]{xy}
\DeclareMathAlphabet{\mathpzc}{OT1}{pzc}{m}{it}

\def\mymathhyphen{{\hbox{-}}}

\mathchardef\myhyphen="2D

\newcounter{example-counter}
\newenvironment{example}%
{\vskip \abovedisplayskip \refstepcounter{example-counter}%
\noindent {\bf Example \arabic{example-counter}.}}%

\newcounter{definition-counter}
\newenvironment{definition}%
{\vskip \abovedisplayskip \refstepcounter{definition-counter}%
\noindent {\bf Definition \arabic{definition-counter}.} \em }%

\newcounter{remark-counter}
\newenvironment{remark}%
{\vskip \abovedisplayskip \refstepcounter{remark-counter}%
\noindent {\bf Remark \arabic{remark-counter}.}}%

\newcommand{\boxtheorem}{\hfill $\blacksquare$\vspace{1mm}}
\newcommand{\ignore}[1]{}
\newcommand{\nit}[1]{{\it #1}}
\newcommand{\Q}{\mathcal{Q}}

\newcommand{\match}{\mathfrak{m}\!}

\newcommand{\eat}[1]{}
\newcommand{\mc}[1]{\mathcal{ #1}}

\newcommand{\defproof}[2]{{\noindent\bf Proof of #1:\
}#2 \boxtheorem\\ \vspace{2mm}}

\newcommand{\hproof}[1]{{\noindent\bf Proof:\
}#1 \boxtheorem\\ }

\newcommand{\red}[1]{#1}
\newcommand{\re}[1]{#1}
\newcommand{\blue}[1]{#1}
\newcommand{\bl}[1]{#1}

\newcommand{\comlb}[1]{{\vspace{2mm}\noindent \bf \blue{COMM(LEO):}}~ #1 \hfill {\bf
    END.}\\}

\newcommand{\comzb}[1]{{\vspace{2mm}\noindent \bf \red{\newline COMM(Zeinab):}}~  #1 \hfill {\bf
    END.}\newline \\}

\newcommand{\comRW}[1]{{\vspace{2mm}\noindent \bf \red{COMM(Reviewer):}}~ #1 \hfill {\bf
    END.}\\}

\newcommand{\ent}[1]{{\sf \small #1}}

\newcommand{\mf}[1]{\mathbf{#1}}

\begin{document}

\begin{frontmatter}

\title{\bf {\em ERBlox}: \ Combining  Matching Dependencies with Machine Learning for Entity Resolution}

\author[mymainaddress]{{\bf Zeinab Bahmani}}

\author[mymainaddress]{{\bf Leopoldo Bertossi}\corref{mycorrespondingauthor}}
\cortext[mycorrespondingauthor]{Corresponding author}
\ead{bertossi@scs.carleton.ca}

\address[mymainaddress]{Carleton University,  School of Computer Science,  Ottawa,  Canada}

\author[mysecondaryaddress]{{\bf Nikolaos Vasiloglou}}
\address[mysecondaryaddress]{LogicBlox Inc.,  Atlanta,  GA 30309,  USA}




\begin{abstract}
Entity resolution (ER), an important and common data cleaning problem, is about
detecting data duplicate representations for the same external entities,
and merging them into single  representations. Relatively recently, declarative rules called
{\em matching dependencies} (MDs) have been proposed for specifying similarity conditions
under which attribute values in database records are merged.  In this work we show the process and the benefits of
integrating four components of ER:  (a) Building a classifier for duplicate/non-duplicate record pairs built
using machine learning (ML) techniques; \ (b) Use of MDs for supporting the blocking phase of ML; \ (c)
Record merging on the basis of the classifier results; and (d) The use of the declarative language {\em LogiQL} -an extended form of Datalog
supported by the {\em LogicBlox} platform- for all activities related to data processing, and the specification and
enforcement of MDs.
\end{abstract}


\end{frontmatter}

\section{Introduction}\label{intro}

\vspace{-2mm}
Entity resolution (ER) is a common and difficult problem in data cleaning that has to do with handling
unintended multiple representations in a database of the same external objects. This problem is also known as deduplication, reference reconciliation, merge-purge, etc. Multiple representations
lead to uncertainty in data and the problem of managing it. Cleaning the database
reduces uncertainty. In more precise terms, ER
is about the identification  and fusion of  database records (think of rows or tuples in tables) that represent the same real-world
entity \cite{naumannACMCS,elmargamid}. As a consequence, ER usually goes through two main consecutive phases: (a) detecting duplicates, and (b) merging them into single representations.

\ignore{\paragraph{\bf Duplicate detection.}} For duplicate detection, one must first analyze multiple pairs of records, comparing the two records in them, and discriminating between: {\em pairs of duplicate records} and {\em pairs of non-duplicate records}. This classification problem is
approached with machine learning (ML)  methods,  to learn  from previously known or already made classifications (a training set for supervised learning), building a {\em classification model} (a classifier) for deciding about other record pairs \cite{Christen2007,elmargamid}.

In principle, in ER every two records (forming a pair) have to be compared through the classifier.
Most of the work on applying ML to ER work at the record level \cite{Rastgoi11, Christen2007, Christen2008}, and only some of the attributes, or their features, i.e.~numerical values associated to them, may be involved in duplicate detection. The choice of relevant sets of attributes and features is application
dependent.

With a classifier at hand, ER may be a task of quadratic complexity since it requires comparing every two records.  To reduce the large number of two-record comparisons, {\em blocking techniques}
are used \cite{surveyBlocking,Baxter03,Herzog07,Garcia-Molina09}. Commonly, a single record attribute, or a combination of attributes, the so-called {\em blocking key}, is used to split the database records into blocks. Next, under the assumption that any two records in different blocks are unlikely to be duplicates, only every two records in a same block are compared for duplicate detection. \ignore{For example, we might block a set of employee records according to the city. We then only need to compare the employees with the same city to detect duplicate employees. Employee pairs that are in different blocks, with different cities, are considered as non-duplicate employee pairs.}

Although blocking will discard many record pairs that are obvious non-duplicates, some true duplicate pairs might  be missed (by putting them in different blocks),
due to errors or typographical
variations in attribute values or the rigidity and low sensitivity of blocking keys. More interestingly, similarity between blocking key values alone may fail to capture the
relationships that naturally hold in the data and could  be used for blocking. Thus, entity blocking  based only on similarities of blocking key values may cause low recall.
This is a major drawback of traditional blocking techniques.

\bl{In this work we consider  different and coexisting entities, for example \ent{Author} and \ent{Paper}. For each of them, there is a collection of records.  For entity \ent{Author},  records may have the form
$\mathbf{a} = \langle \nit{name}, \ldots, \nit{affiliation}, \ldots,$ $\nit{paper~title}, \ldots\rangle$; and for \ent{Paper} entity, records may be of the form
$\mathbf{p} = \langle \nit{title}, \ldots, \nit{author~name}, \ldots\rangle$.\footnote{For all practical purposes, think of records as database tuples in a single table.}}

Records for different entities may be related
via attributes in common and referential constraints, something the blocking mechanism could take advantage of. Blocking can be performed on each of the participating entities, and the way records \bl{for an entity, say \ent{Author}, are placed in blocks
may influence the way the records for another entity, say \ent{Paper}, are assigned to blocks.} This is called ``collective blocking". Semantic, {\em relational} information, in addition to that provided
by blocking keys for single entities, can be used to
state relationships between different entities and their corresponding similarity criteria. So, blocking decision making forms a  collective and intertwined process involving several entities. In the end,
the records for each individual entity will be placed in blocks associated to that entity.

\ignore{\red{Collective blocking is a natural extension
of the notion of blocking in which one wants to block different types
of real-world entities in a set of records at the same time, in an intertwined process. In other words, in collective blocking, the semantic knowledge is used to make all the blocking decisions collectively, in addition to blocking keys which are used to block a single entity type. In general, the output of collective blocking is sets of blocks of the input records (by entity type).} }

 In our work, collective blocking is based on blocking keys and {\em the enforcement} of semantic information about the {\em relational closeness} of entities \ent{Author}
and \ent{Paper}, which is captured by a set of {\em matching dependencies} (MDs) \cite{FanJLM09}. So, we propose ``MD-based collective blocking".

After records are divided in blocks, the proper duplicate detection process starts, and is carried out by comparing every two records
in a block, and classifying the pair as ``duplicates" or ``non-duplicates" using the trained ML model at hand. In the end,  records in duplicate pairs are
 considered to represent the same external entity, and have to be {\em merged} into a single representation,
i.e.~into a single record. This second phase is also application dependent. MDs were originally proposed to support this kind of task, and their use in blocking is somehow unexpected.

Matching dependencies are declarative logical rules that tell us under what conditions of similarity between attribute values, any two records must have certain attribute values merged (or matched), i.e.~made identical
\cite{Fan08,FanJLM09}. For example, the MD: \vspace{-2mm}
\begin{equation}\nit{Dept}_{\!B}[\nit{Dep}] \approx \nit{Dept}_{\!B}[\nit{Dep}] \ \to \ \nit{Dept}_{\!B}[\nit{City}] \doteq \nit{Dept}_{\!B}[\nit{City}] \label{eq:md}
\end{equation}
tells us that, for any two records for entity (or relation or table) $\nit{Dept}_{\!B}$ that have similar values for attribute $\nit{Dep}$,  their values
for attribute $\nit{City}$ should be merged, i.e.~made the same.

MDs as introduced in \cite{FanJLM09} do not specify how to merge values. In \cite{Bertossi12}, MDs were extended with {\em matching functions} (MFs). For a  data domain, a MF
specifies how to assign a value in common to two values. In this work, we adopt MDs with MFs. In the end, the enforcement of MDs with MFs should produce a duplicate-free instance
(cf.~Section \ref{pre} for more details).

MDs have to be specified in a declarative manner, and at some point enforced, by producing changes on the data. For this purpose, we use the {\em LogicBlox} platform, a data management system developed by the LogicBlox\footnote{ www.logicblox.com} company, that is centered around its declarative language, {\em LogiQL} \cite{Halpin15}. {\em LogiQL} supports relational data management and, among several other features  \cite{Aref15},  an extended form of Datalog with stratified negation \cite{ceri90}. This language is expressive enough for the kind of MDs considered in this work.\footnote{For arbitrary sets of MDs, we need higher expressive power \cite{Bertossi12}, such as that provided by {\em answer
set programming} \cite{Bahmani12}.}

In this paper, we describe our {\em ERBlox} system. It is built on top of the {\em LogicBlox} platform, and implements entity resolution (ER) applying {\em LogiQL} for the specification and enforcement
of MDs, and built-in  ML techniques for building the classifier.
More specifically, {\em ERBlox} has \red{four main components or modules}:
\red{\begin{itemize}
\item[(a)] {\em MD-based collective blocking:} \  This phase is just about clustering together records that might be duplicates of each other. Additional comparisons between two records between will be performed within block, and never with records from different blocks. Blocking can be used before learning a classifier, to ascribe labels (duplicate/non-duplicate or $\pm1$) to pairs of records that will become training examples, or after the classifier has been learned, with new records in the database that have to be checked for duplication with other records in the database.\footnote{\red{In our case, the training data already came with labels. So, blocking was applied to the unlabeled records before, but independently, from the learning and execution of the classifier.}}
     It may be the case that two records in a same block may end up not being considered as duplicates of each other. However, through blocking the number of pairwise record comparisons of is reduced.
 \item [(b)] {\em ML-based classification model construction:} \ At this point any supervised technique for classification, i.e.~for  building the mathematical model for classification, could be used. This is the proper machine learning phase. We used the {\em support-vector machine} (SVM) approach \cite{cristianini,Vapnik98}.
 \item [(c)] {\em Duplicate detection:} \ Having the new records in the database (as opposed to training examples) already clustered in blocks, this phase is about applying the classification model obtained in the previous phase to new pairs of records, obtaining for each pair the outcome $\pm1$. The classifier could be applied to any two records, or -if blocking techniques have been applied to the database- only to two records in a same block. In our case, we did the latter.
 \item
[(d)] {\em MD-based duplicate merging:} \ The output of the preceding phase is just a set of record-pairs with their newly created labels, $\pm1$, indicating that they are duplicates of each other, or not. This last phase merges duplicates into single records. In our case, when and how to merge is specified by matching dependencies, which invoke  matching functions to find the values in common to be used  in the records created by merging.
\end{itemize}}

The blocking phase, (a) above, uses  MDs to specify the blocking strategy. They express conditions in terms of blocking key similarities and also relational closeness -the semantic knowledge- to assign two records to a same block, by making their block identifiers identical.
Then, under MD-based collective blocking different records of possibly several related entities are simultaneously assigned to blocks through
the enforcement of MDs {(cf.~Section \ref{MDBlocking} for details). This is a non-traditional, novel use of MDs, whereas their intended use is the application to proper merging phase, (d) above, \cite{Fan08}.

\red{It is important to emphasize that, in our work, MDs {\em  are not} used for the whole ER process, but only in two of the phases above. In principle, the whole ER process could be based only on MDs. However, this would be a complete different problem, in particular, a completely different machine learning problem: the MDs for this application would have to be learned from scratch (implicitly learning similarity relationships and a classifier). Learning MDs is a rather unexplored area of research (cf.~\cite{leiChen,leiChen2} for some  work in this direction), which is somehow closer to the areas of {\em rule learning} \cite{fur+} and  discovery of database dependencies \cite{felix}.  With our approach we can leverage, at a particular phase of the ER process, available machine learning techniques that are fully integrated with database management systems, as in the case of {\em LogicBlox}.}

The sets of MDs used in (a) and (d) are different, and play different roles. In both cases, they are application-dependent, and have a canonical  representation in the system, as Datalog rules.
The MDs are then enforced by applying (running) those rules. Although in general a  set of MDs may lead to alternative final instances  through its enforcement \cite{Bertossi12}, in our application of MDs both sets of MDs lead to a single instance.

 In the case of (a), this means that, for each entity, a unique set of disjoint blocks  is generated. The reason is that the combination of the set of MDs and the initial database instance falls into a newly identified, well-behaved
  class, the {\em SFAI class}, that we introduce in this work. (The main ideas and intuitions around it are presented in the main body of this paper, but more specific details are given in \ref{sec:relMDs}.) \  In the case of
 (d), the set of ``merge" MDs also leads to  a single,
duplicate-free instance (as captured by the classifier and the merge MDs). This is because the MDs in the set turn out to be  {\em interaction-free} \cite{Bertossi12}(cf.~also \ref{sec:relMDs}).

We use {\em LogiQL} to declaratively implement the two MD-based
components of {\em ERBlox}. As shown in \cite{Bahmani12,BahmaniExten12} in general, sets of MDs can be expressed by means of {\em answer-set programs} (ASPs) \cite{brewka}. However, both classes of MDs used by {\em ERBlox} can be expressed by computationally efficient fragments of ASPs, namely Datalog with stratified negation \cite{ceri90}, which is supported by {\em LogiQL}.

On the machine learning side (item (b) above), the problem is about building and implementing a model for the detection of pairs of duplicate records. The classification model is trained using record-pairs known to be duplicates or non-duplicates. We independently used three established classification algorithms: SVM, {\em k-nearest neighbor} (K-NN) \cite{Cover67}, and {\em non-parametric Bayes classifier} (NBC) \cite{Baudat00}. We used the {\em Ismion}\footnote{http://www.ismion.com} implementation of them
due to the in-house expertise at LogicBlox. Since the emphasis of this work is on the use of {\em LogiQL} and MDs, we will refer only to our use of SVM.


For experimentation  with the {\em ERBlox} system, we  used as dataset  a snapshot of Microsoft Academic Search (MAS)\footnote{http://academic.research.microsoft.com. As of January 2013.} that includes $250$K authors, $2.5$M papers, and a training set. We also used, independently, datasets from DBLP
and Cora Citation. The experimental results show that our system improves ER recall and precision over traditional, {\em standard}  blocking techniques \cite{jaro89},  where just blocking-key similarities are used. Actually, MD-based collective blocking leads to higher precision and recall on the given datasets.  \ignore{For comparison, we also tested our system with data from DBLP and Cora.}

Our work also shows the integration under a single system of different forms of data retrieval, storage and transformation, on one side, and machine learning techniques, on the other. All this is
enabled by the use of optimized Datalog-rule declaration and execution as supported by the {\em LogicBlox} platform.

\bl{This paper is structured as follows. Section \ref{pre} introduces background on: \ matching dependencies (including a brief description of the new SFAI class), classification, and collective blocking. A general overview of the {\em ERBlox} system is presented in Section \ref{MLMDFramework}. Specific details
about the components of our methodology and {\em ERBlox} are given and discussed in Sections \ref{sec:init}, \ref{MDBlocking}, \ref{sec:MLClassification}, and \ref{Detection}. Experimental
 results are shown in Section \ref{evaluation}. Sections \ref{RelWork} and \ref{conclude} present related work and conclusions, respectively. \ignore{\ref{sec:extExample} contains an extended example
 illustrating the properties of blocking-MDs.} In \ref{sec:relMDs} we provide the definitions and more details about relational MDs, the SFAI class, and other classes with the {\em unique clean instance} property.\footnote{The material in \ref{sec:relMDs} is all new, but, although important for ERBlox, departs from the main thread of the paper.} This paper is a revised and extended version of \cite{sum}.}

\section{Preliminaries}\label{pre}

\subsection{Matching dependencies}\label{sec:mds}


We consider an application-dependent relational schema $\mc{R}$, with a data domain $U$.
For an attribute $A$, $\nit{Dom}(A) \subseteq U$ is its domain. We assume predicates
do not share attributes, but different attributes may share a domain.
An instance $D$ for $\mc{R}$ is a finite set of ground atoms of the form $R(c_1,\ldots, c_n)$, with $R \in \mc{R}$,
$c_i \in U$. The {\em active domain} of an instance $D$, denoted $\nit{Adom}(D)$, is the finite set of all constants from $U$ that appear in $D$.

We assume that each entity is represented by a relational predicate, and its tuples or rows
in its extension correspond to records for the entity. As in \cite{Bertossi12}, we assume records
have unique, fixed, \bl{{\em global record identifiers} ({\em rids})}, which are positive integers. This allows us to  trace changes of attribute values in records.
\bl{When records are represented as tuples in a database, which is usually the case, we talk about {\em global tuple identifiers} ({\em tids}).}
\bl{Record and tuple ids are placed in an extra, first attribute} for $R \in \mc{R}$ that acts as a key. \bl{Then, records take the form $R(t,\bar{c})$, with $t$ the identifier.
Sometimes we leave tids and rids implicit. \ If $\mc{A}$ is a sublist of the attributes for a predicate $R$, $R[\mc{A}]$ denotes the restriction of an $R$-tuple (or the predicate $R$) to attributes
in $\mc{A}$.}


MDs  are formulas of the form \cite{Fan08,FanJLM09}:

\begin{equation}
\varphi\!:  \ \ \bigwedge_j R_1[X_1^j] \approx_j R_2[X_2^j] \ \longrightarrow \ \bigwedge_k R_1[Y_1^k] \doteq R_2[Y_2^k],\label{eq:md2145}
\end{equation}
\vspace{-6mm}
\phantom{oo}

\noindent
 where  \bl{attributes (treated as variables)} $X_1^j$ and  $X_2^j$ (also  $Y_1^k,Y_2^k$) are
{\em comparable},  in the sense that they share
 the same data domain $\nit{Dom}_j$ on which a binary similarity (i.e., reflexive and symmetric) relation $\approx_j$ is defined. $R_1, R_2$ could be the same predicate. The MD in (\ref{eq:md2145}) states that, for every pair of tuples (one in relation $R_1$, the other in relation $R_2$) where the left-hand side (LHS) of the arrow is true, the attribute values in them on the right-hand side (RHS) have to be made identical. We can consider only MDs with a single identity atom (with $\doteq$) in the RHSs. \bl{Accordingly, an explicit formulation of the MD in (\ref{eq:md2145}) in classical predicate logic is:\footnote{\bl{Similarity symbols can be treated as regular, built-in, binary predicates, but
 the identity symbol, $\doteq$, would be non-classical.}}}
\begin{equation}
\bl{\varphi\!:  \ \ \forall t_1 t_2 \ \forall \bar{x}_1 \bar{x}_2(R_1(t_1,  \bar{x}_1) \wedge R_2(t_2,  \bar{x}_2) \  \wedge \ \bigwedge_j x_1^j \approx_j x_2^j \ \  \longrightarrow \ \ y_1 \doteq y_2),}\label{eq:md12145}
\end{equation}
\vspace{-6mm}
\phantom{oo}

\noindent \bl{with $x_1^j, y_1 \in \bar{x}_1, \ x_2^j, y_2\in \bar{x}_2$.  The $t_i$ are used as variables for  tuple IDs. We usually leave the universal quantifiers implicit. \ignore{In (\ref{eq:md12145}), $\approx_j$ is a binary similarity relation  on domain $\nit{Dom}_j$.}  $\nit{LHS}(\varphi)$ and $\nit{RHS}(\varphi)$ denote the sets of
atoms on the LHS and RHS of $\varphi$,
respectively. \ $\nit{LHS}(\varphi)$ contains, apart from similarity  atoms, atoms $R_1(t_1, \bar{x}_1)$ and $R_2(t_2, \bar{x}_2)$, which contain all the variables in the MD, including those in the $\nit{RHS}(\varphi)$.  \ignore{such that some of the variables in $\bar{x}_1, \bar{x}_2$ are used in similarity atoms and equalities in $\varphi_1$.} So,   similarity  and identity atoms  in $\varphi$ involve  one  variable from predicate $R_1$, and one from predicate $R_2$.}

\bl{\begin{example} \label{ex:first} Consider $\nit{Paper(PID,Title,Year,CID,JID,Keyword,Bl \#})$, a relational predicate  representing records for entity \ent{Paper}. It includes a first attribute for a tuple identifier, and a last indicating the block the tuple (record) has been assigned to. The MD
\begin{eqnarray}
\nit{Paper}(\nit{pid}_1,x_1,y_1,z_1,w_1,v_1,\nit{bl}_1) \ \wedge\nit{Paper}(\nit{pid}_2,x_2,y_2,z_2,w_2,v_2,\nit{bl}_2) \ \wedge \nonumber\\
x_1 \approx_{\!\nit{Title}} x_2 \ \wedge \ y_1=y_2 \ \wedge z_1=z_2 \ \longrightarrow \ \nit{bl}_1 \doteq \nit{bl}_2, \label{eq:block}
\end{eqnarray}
involves a similarity relation on the \nit{Title} attribute, and equality as similarity relation on attributes \nit{Year} and \nit{CID}. The MD specifies that, when the conditions
expressed in the LHS are satisfied, the two block values have to be made the same, i.e.~the two records should be (re)assigned to the same block. \boxtheorem
\end{example}}

 A {\em dynamic, chase-based semantics} for MDs with {\em matching functions} (MFs) was introduced in \cite{Bertossi12}, and we briefly summarize it here.  Given an initial instance $D$, the set $\Sigma$ of MDs is iteratively enforced until they cannot be be applied any further, at which point
a {\em resolved instance} has been produced.

In order to {\em enforce} (the RHSs of) MDs,  there are binary {\em matching functions} (MFs)
$\match_A: {\it Dom}(A)\times {\it Dom}(A)\rightarrow {\it Dom}(A)$; and
$\match_A(a,a')$ is used to replace two values $a, a' \in {\it Dom}(A)$ that have to be made identical.  For example,
for an attribute $\nit{Address}$, we might have a MF $\match_{\it Address}$, such that $\match_{\it Address}(\mbox{``MainSt., Ottawa"}, \mbox{``25 Main St."}) := \mbox{``25 MainSt., Ottawa"}$.

MFs are idempotent, commutative, and associative, and then induce a partial-order
structure $\langle {\it Dom}(A), \preceq_A\rangle$, with: $a \preceq_A a' \ :\Leftrightarrow \ \match_A(a,a') = a'$ \ \cite{Bertossi12,BenjellounGMSWW09}.
It always holds: \ $a,a' \ \preceq_A \ \match_A(a,a')$. Actually, the relationship $a \preceq_A a'$ can be thought in terms of {\em information contents}: \
$a'$ is at least as informative as $a$.\footnote{\red{Of course, this claim assumes that MFs locally assign an at least as informative value as both of the two input values. MFs are application dependent.}}  This partial order allows to define a partial order  $\sqsubseteq$ on instances \cite{Bertossi12}. Accordingly, when
MDs are applied, a chain of increasingly more informative (or less uncertain) instances is generated: \ $D_0 \sqsubseteq D_1  \sqsubseteq \cdots \sqsubseteq D_{\it clean}$.
  In this work, MFs are treated as built-in relations.

Given a database instance $D$ and a set of MDs $\Sigma$, there  may be several resolved instances for $D$ and $\Sigma$ \cite{Bertossi12}.
However, \bl{there is a unique resolved instance  if one of the following holds} \cite{Bertossi12,BahmaniExten12}:
\begin{itemize}
\item[(a)] MFs used by $\Sigma$ are {\em similarity-preserving}, i.e., for every $a, a', a''\!: \  \ a \approx a' \mbox{ implies } a \approx \match_A(a',a'')$. \ \bl{When MDs use similarity-preserving MFs, we also say that the
MDs are similarity-preserving.}

\item[(b)] $\Sigma$ is {\em interaction-free}, i.e.~no attribute (with its predicate) appears both in a RHS and  a LHS of MDs in $\Sigma$.

For example, the set $\Sigma_1 = \{R[A] \approx T[B] \rightarrow R[C] \doteq T[D], \ \  T[D] \approx S[A] \rightarrow T[A] \doteq S[B]\}$ is not interaction-free due to the presence
of attribute $T[D]$. \ $\Sigma_2 = \{R[A] \approx T[B] \rightarrow R[C] \doteq T[D], \ \  T[A] \approx S[A] \rightarrow T[C] \doteq S[C]\}$ is interaction-free.

\item[(c)] The {\em combination} of $\Sigma$ and the initial instance $D$ is {\em similarity-free attribute intersection} (we say it is SFAI), if $\Sigma$ is interaction-free, or,  otherwise, for every pair of interacting MDs $\varphi_1, \varphi_2$ in $\Sigma$, and, for every $t_1,t_2, t_3 \in D$,  it holds   $\nit{LHS}(\varphi_1)$ is not true in instance $\{t_1, t_2\}$ or $\nit{LHS}(\varphi_2)$ is not true in instance $\{t_2,t_3\}$.

Consider, for example, predicate $R(A,B,C)$, the instance $D$ below,  and the set $\Sigma$ of interacting MDs:
\begin{eqnarray*}
\varphi_1\!:&& \ R\left[A\right] \approx R\left[A\right] \ \longrightarrow \ R\left[B\right] \doteq R\left[B\right],\\
\varphi_2\!:&& \ R\left[B\right] \approx R\left[B\right] \ \longrightarrow \ R\left[C\right] \doteq R\left[C\right].
\end{eqnarray*}
\begin{multicols}{2}
\begin{center}
\begin{tabular}{c|c|c|c|}\hline
$R(D)$&$A$ & $B$ & $C$\\ \hline
$t_1$&$a_1$ & $b_1$ & $c_1$\\
$t_2$&$a_2$ & $b_2$ & $c_2$\\
$t_3$&$a_3$ & $b_3$ & $c_3$\\ \cline{2-4}
\end{tabular}
\end{center}
Assume that the only similarities that holds in the data domain $U$ are $a_1\approx_A a_2$ and $b_1 \approx_B b_4$, with $b_4 \in \nit{Dom}(B) \smallsetminus \nit{Adom}(D)$.
\end{multicols}
 Since $\varphi_2$ is not applicable in $D$ (i.e.~there is no pair of tuples making it true), the combination of $\Sigma$ and $D$ is SFAI. Notice that $b_1 \approx_B b_4$ does not matter, because there is no tuple in $D$ with $b_4$ as value for $R[B]$.

\bl{With general sets of MDs, different orders of MD enforcements  may result in different clean instances, because tuple similarities may be broken  during the chase with interacting, non-similarity-preserving MDs, without reappearing again \cite{Bertossi12}.
With SFAI combinations, two similar tuples in the original instance $D$ -or becoming similar along a chase sequence- may have the similarities broken
in a chase sequence, but they will   reappear later on in the same and the other chase sequences. Thus, different orders of MD enforcements cannot lead in the end to different clean instances.}

\end{itemize}

\bl{ The SFAI class had not been investigated before. It is a semantic class, as opposed to syntactic, in that there is a dependency upon the initial instance. See \ref{sec:relMDs} for more details on this class.}

\bl{The three classes above have the {\em unique clean instance} (UCI) property, i.e.~iteratively and exhaustively enforcing them leads to a single clean, stable instance. } \red{Even more, in these three cases, the single clean instance can be computed
in polynomial time {\em in data}, i.e.~in polynomial time in  $|D|$, the size of the initial instance, leaving the set of MDs as a fixed, external parameter for the computational problem that here receives database instances as inputs.}\footnote{\re{In data management it is common to measure computational complexity (in our case, time complexity) in terms of the size of the underlying dataset, which is usually much larger than that of any other ingredient, such as a query, a set of integrity constrains, a set of view definitions, etc. If we bring the sizes of the latter into the complexity analysis, we talk of {\em combined complexity} \cite{ahv95}.}}

\bl{In this work, for collective-blocking purposes,  we will introduce and use a new class of MDs, that of {\em relational MDs}, that extends the class of ``classical" MDs introduced earlier
in this section. Actually, the three UCI classes of classical MDs listed above can be extended to relational MDs, and preserving the UCI property (cf.~\ref{sec:relMDs} for more details).}

\bl{Relational MDs, the SFAI class, and the UCI property are all relevant for this work. However, a detailed analysis of them is somehow beyond the scope of this work. For this reason, and in order not to break the natural flow of the presentation,  we provide in \ref{sec:relMDs}, mainly for reference,
some more details about all these subjects.}

\subsection{Classification with support-vector models}

The {\em support-vector machines} \ technique (SVM)  \cite{Vapnik98} is a form of  kernel-based learning.  \ SVM can be used for classifying vectors in  an inner-product vector space $\mc{V}$
over $\mathbb{R}$. Vectors are classified
 in two classes, say with labels $0$ or $1$. The classification model is a hyper-plane in $\mc{V}$: vectors are classified depending on the side of
 the hyperplane they fall.

 The hyper-plane has to be learned through an algorithm applied to a training set of examples, say  $E = \{(\mathbf{e}_1, f(\mathbf{e}_1)), (\mathbf{e}_2,$ $f(\mathbf{e}_2)), (\mathbf{e}_3, f(\mathbf{e}_3)),$ $\ldots, (\mathbf{e}_n,$ $f(\mathbf{e}_n))\}$. Here,
  $\mf{e}_i \in \mc{V}$, and for the real-valued {\em feature} (function) $f$: \ $f(\mf{e}_i) \in \{0, 1\}$.

\begin{figure}[h]
\hspace*{4.5cm}\includegraphics[width=5.5cm]{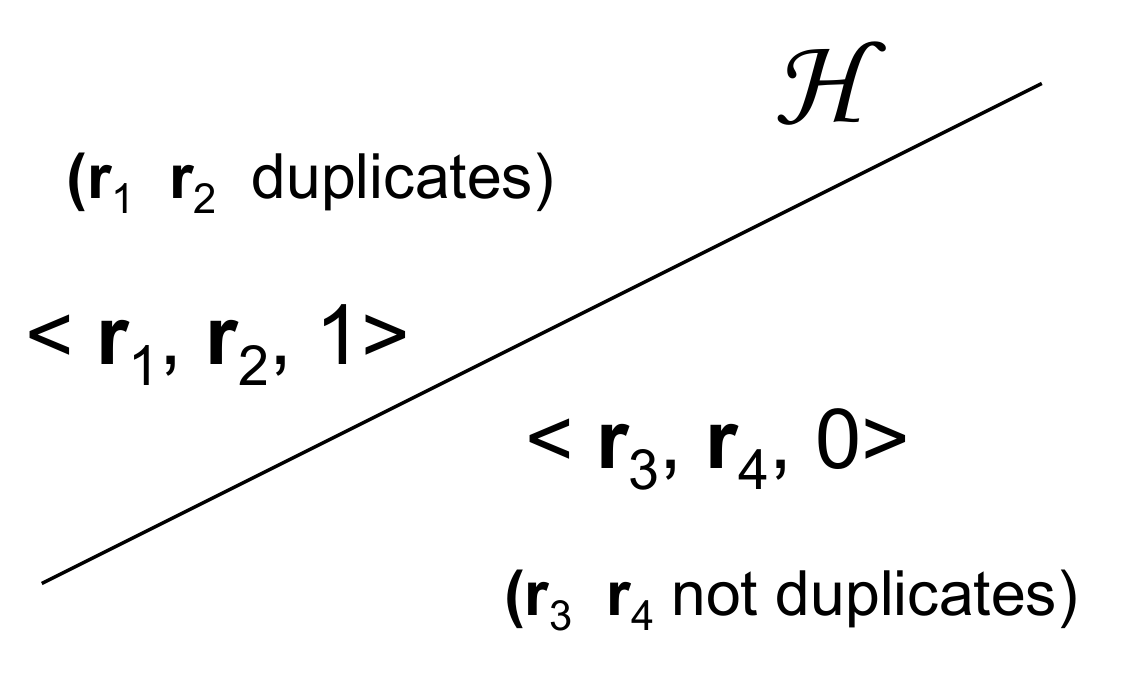}
\vspace{-6mm}
\caption{~~Classification hyperplane}\label{fig:sep} \vspace{-3mm}
\end{figure}

The SVM algorithm  finds an optimal hyperplane, $\mc{H}$, in $\mc{V}$ that  separates the  two classes in which the training vectors are classified. Hyperplane $\mc{H}$ has an equation of the form $\mathbf{w} \bullet  \mathbf{x} + b$, where
$\bullet$ denotes the inner product, $\mf{x}$ is a vector variable, $\mf{w}$ is a weight-vector of real values, and $b$ is a real number.
Now, a new vector $\mf{e}$ in $\mc{V}$ can be classified as positive or negative depending on the side of $\mc{H}$ it lies. This is determined by computing $h(\mathbf{e}) := \nit{sign}(\mathbf{w} \bullet  \mathbf{e} + b)$. 
If $h(\mathbf{e})> 0$, $\mathbf{e}$ belongs to class $1$; otherwise, to class $0$.

It is possible to compute real numbers $\alpha_1, \ldots, \alpha_n$, the coefficients of the ``support vectors", such that the classifier $h$ can be computed through:  $h(\mathbf{e}) = \nit{sign}(\sum_i \alpha_i \cdot f(\mathbf{e}_i) \cdot \mathbf{e}_i \bullet \mathbf{e} + b)$ \ \cite{flach}.

As Figure~\ref{fig:sep} shows, in our case, we need to classify {\em pairs of records}, that is our vectors are of {\em record-pairs} of the form $\mathbf{e} = \langle r_1, r_2\rangle$. If   $h(\mathbf{e}) = 1$, the classifier returns as output $\langle r_1, r_2, 1\rangle$, meaning that they two records are duplicates (of each other). Otherwise, it returns
$\langle r_1, r_2, 0\rangle$, meaning that the records are non-duplicates (of each other). For the moment we do not need more than this about the SVM technique.


\subsection{\red{Collective blocking}} \label{sec:colBl}

\red{Entity-resolution (and other machine learning tasks) use {blocking techniques}, to group together input values for further processing. In the case
of ER, records that might be duplicates of each other are grouped under a same block, and only records within the same block  are compared. Any two records in different blocks will never be  declared as duplicates.}

\red{ Commonly,  a single  attribute in records, or a combination of attributes,
called a  {\em blocking key}, is used to split records into blocks. If two records share the same (or have similar) values for the blocking-key attributes, they are put into the same block.
 For example, we could block employee records according to the name and  the city. If two of them share (or have similar) name and city values, they go to the same block. Additional analysis, or the use of a classifier, will eventually determine if they are duplicates or not.}

\red{Blocking keys are rather rigid, and ``local", in that they are applied to records for a single entity (other entities may have other blocking keys). Their use may cause low recall. For this reason, it may be  useful to apply blocking techniques that take advantage of {\em additional
semantics} and/or {\em domain knowledge}. Actually, collective blocking  creates blocks for different entities by exploiting the relational relationships
between entities. Records for different entities are separately, but simultaneously blocked, in interaction. Accordingly, this approach can be called {\em semantic collective blocking}.}\\

\begin{figure}[h]
\begin{center}
\begin{center}
\includegraphics[width=7.5cm]{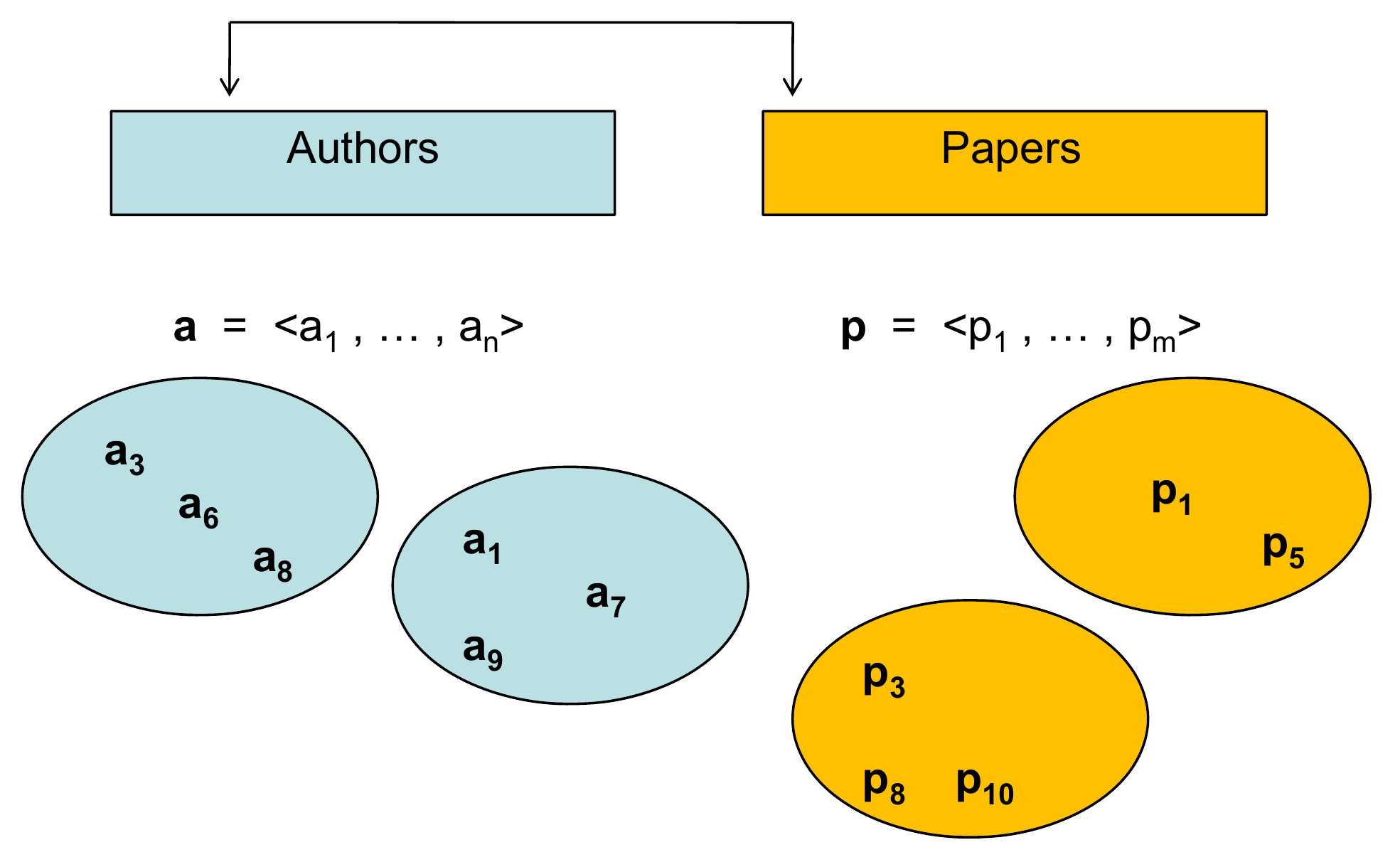}
\end{center}
\vspace{-4mm}
\caption{\ \ Collective blocking}\label{fig:blocks}
\vspace{-6mm}
\end{center}
\end{figure}

\begin{example}
\label{ex:IntroBlock}
\red{Consider two entities, \ent{Author} and \ent{Paper}. For each of them, there is a set of
records. For \ent{Author}, they are  of the form
$\mathbf{a} = \langle \nit{name}, \ldots, \nit{affiliation}, \ldots,$ $\nit{paper~title}, \ldots\rangle$, with \{\nit{name}, \nit{affiliation}\}  the blocking key; and for \ent{Paper}, records are the form
$\mathbf{p} = \langle \nit{title}, \ldots, \nit{author~name}, \ldots\rangle$, with \nit{title} the blocking key.}

\red{We can block together two \ent{Author} records on the basis of the similarities of their values for the blocking key,  in this case of authors' names and  affiliations. (This blocking policy can be specified
by means of an MD of the form (\ref{eq:block}) in Example \ref{ex:first}.) \ However, if two  \ent{Author} records, say $\mathbf{a_2}, \mathbf{a_5}$, have similar  names, but not similar affiliations, they will not be assigned to the same block.}

\red{An alternative approach, could create and assign blocks of \ent{Author} records, and also blocks of {\sf Paper} records, at the same time, separately for each entity, but in an intertwined process.
In this case, the same \ent{Author} records $\mathbf{a_2}, \mathbf{a_5}$, that were assigned to different blocks, may be the authors of  papers, represented as  {\sf Paper} records, say $\mf{p_3}, \mf{p_8}$, resp., which have been already be put in the same block (of papers) on the basis of similarities of paper titles (cf.~Figure~\ref{fig:blocks}). With this additional information, we might assign $\mathbf{a_2}$ and $\mathbf{a_5}$ to the same block. }

\red{The additional knowledge comes in two forms: (a) {\em semantic knowledge}, about the relational relationships between records for different entities, in this case, the reference of paper titles appearing in \ent{Author} records to paper titles in \ent{Paper} entities, and (b) ``procedural" knowledge that tells us about blocks certain entities have been assigned to. As we will see, MDs allow us to express both, simultaneously. In this case, we will be able
to express that ``if two papers are in the same block, then the corresponding \ent{Author} records that have similar author names should be put in the same block too".
 So, we are blocking \ent{Author} and \ent{Paper} entities, separately, but collectively and in interaction. Similarly, and the other way around, we could block \ent{Papers} records according to the blocking results for their
authors (\ent{Author} records).} \boxtheorem
\end{example}}

\section{Overview of {\em ERBlox}}\label{MLMDFramework}

\begin{figure}[t]
\begin{center}
\includegraphics[width=12.5cm]{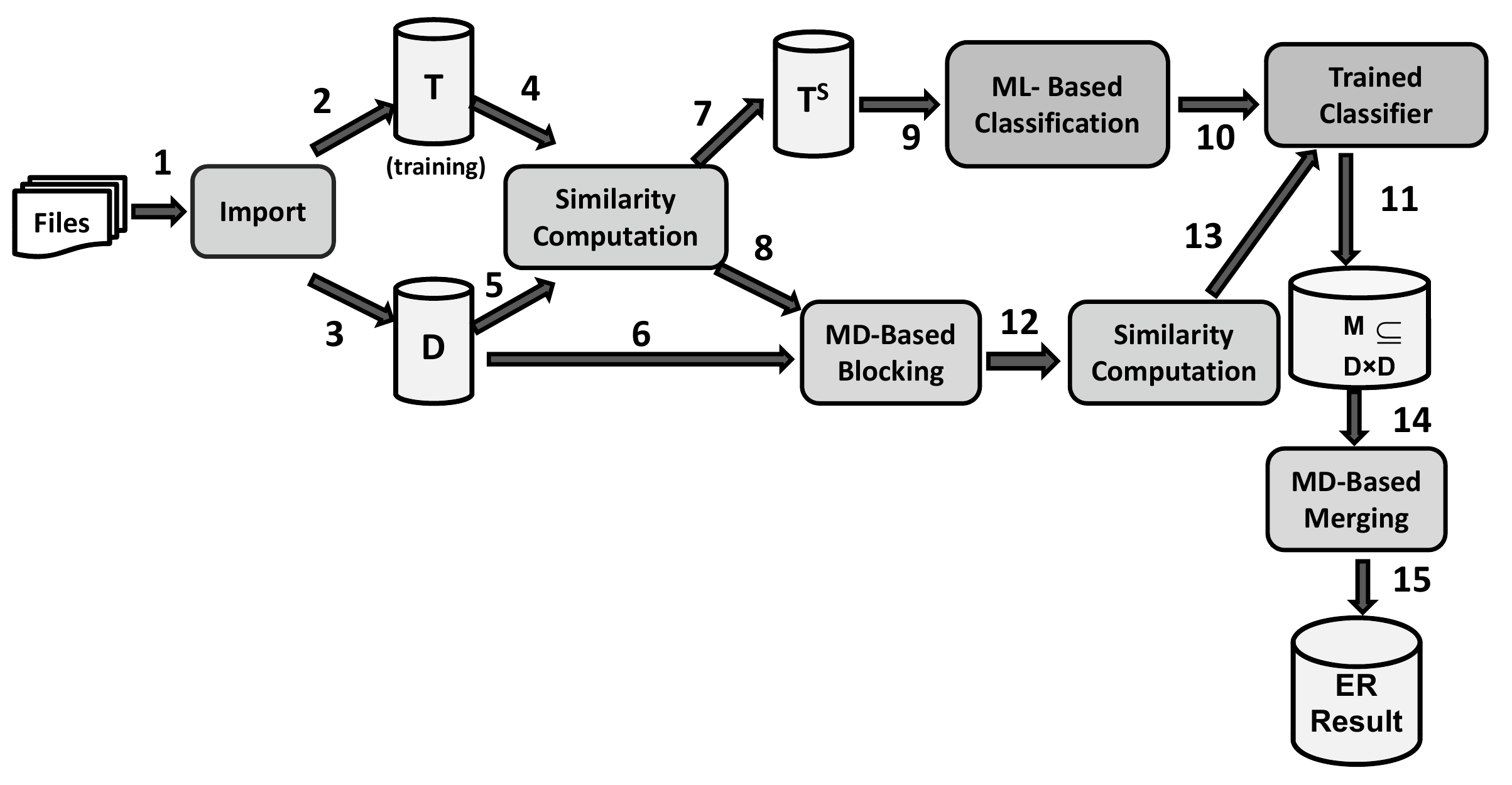}
\vspace*{-1.5cm}
  \caption{\ \ Overview of {\em ERBlox}}\label{fig:ERBlox}\vspace{-6mm}
\end{center}
\end{figure}

A high-level description of the components and workflow of {\em ERBlox} is given in Figure~\ref{fig:ERBlox}. In the rest of this section, numbers in boldface refer to the edges in that figure. {\em ERBlox}'s main four components are: \ 1. \ MD-based collective blocking (path $\mf{1,3,5,\{6,8\}}$), \ 2.  \ Classification-model construction (all the tasks up to $\mf{12}$, inclusive), \ 3. \ Duplicate detection (continues with edge $\mf{13}$), and \ 4. \ MD-based merging (previous path extended with $\mf{14, 15}$).
  All the tasks in the figure, except for the classification model construction (that applying the SVM algorithm), are supported by {\em LogiQL}.\footnote{The implementation of in-house developed ML algorithms as components of the LogicBlox platform is ongoing work.}

The initial input data is stored in structured text files, which are initially standardized and
free of misspellings, etc. However, there may be duplicates.   The general {\em LogiQL} program supporting the above workflow contains rules for importing data from the files into the extensions of relational predicates
(tables). This is  edge $\mf{1}$.  This results in a relational database instance $T$ containing the training data (edge $\mf{2}$), and instance $D$ to be subject to ER (edge $\mf{3}$).

\begin{figure}[h]
\begin{center}
\includegraphics[width=8.5cm]{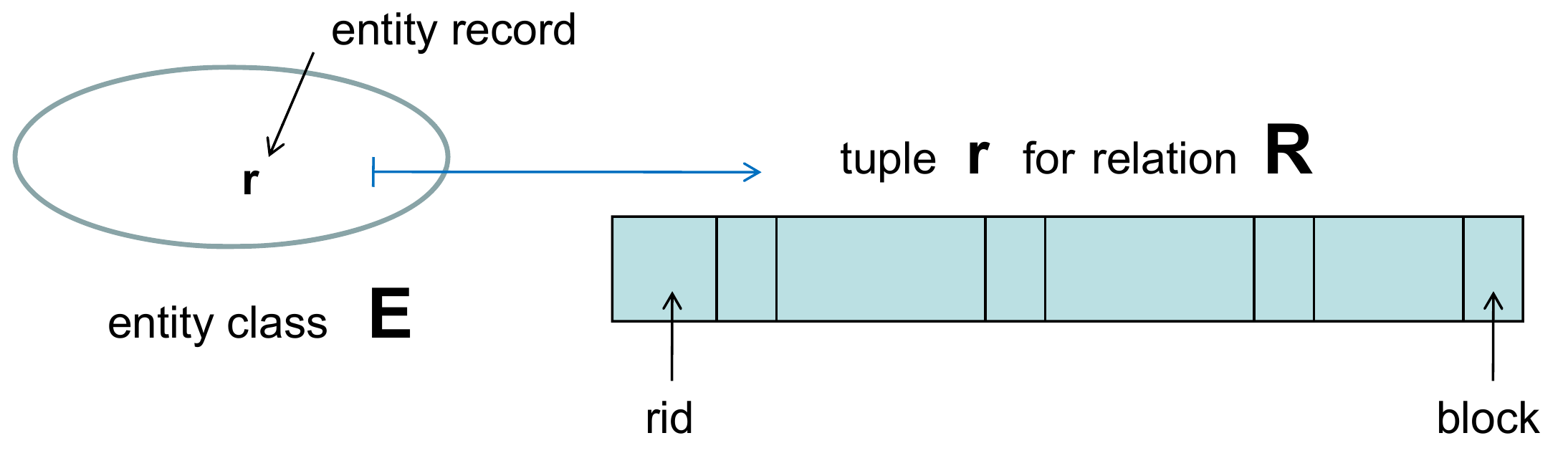}
\vspace*{-0.5cm}
  \caption{\ \ Records}\label{fig:recs}\vspace{-5mm}
\end{center}
\end{figure}

Entity records are represented as relational tuples as shown in Figure~\ref{fig:recs}. However, we will keep referring to them as records, and they will be generally denoted with $r, r_1, ...$.

 The next tasks require similarity computation of pairs of records $\langle r_1, r_2\rangle$ in $T$ and (separately) in $D$ (edges $\mf{4}$ and $\mf{5}$).   Similarity computation is based on two-argument similarity functions on the domain of a record attribute, say \
$\nit{f}_i\!: \nit{Dom}(A_i) \times \nit{Dom}(A_i) \rightarrow [0,1]$, each of which assigns a numerical value to (the comparison of) two values for attribute $A_i$, in two different records.

These similarity functions, being real-valued functions of the objects under classification, correspond to  {\em features} in the general context of machine learning. They are
considered only for a pre-chosen subset of record attributes. Weight-vectors \ $w(r_1,r_2) = \langle \cdots, w_i(\nit{f}_i(r_1[A_i],r_2[A_i])), \cdots \rangle$
are formed by applying predefined weights, $w_i$, on real-valued similarity functions, $f_i$, on a pair of of values for attributes $A_i$ (edges $\mf{4}$ and $\mf{5}$), as in Figure~\ref{fig:sim}. (For more details on similarity computation see Section \ref{sec:init}.)

\ignore{   \begin{wrapfigure}[9]{r}{0.45\textwidth}
\begin{center}
\begin{center}
\includegraphics[width=5.5cm]{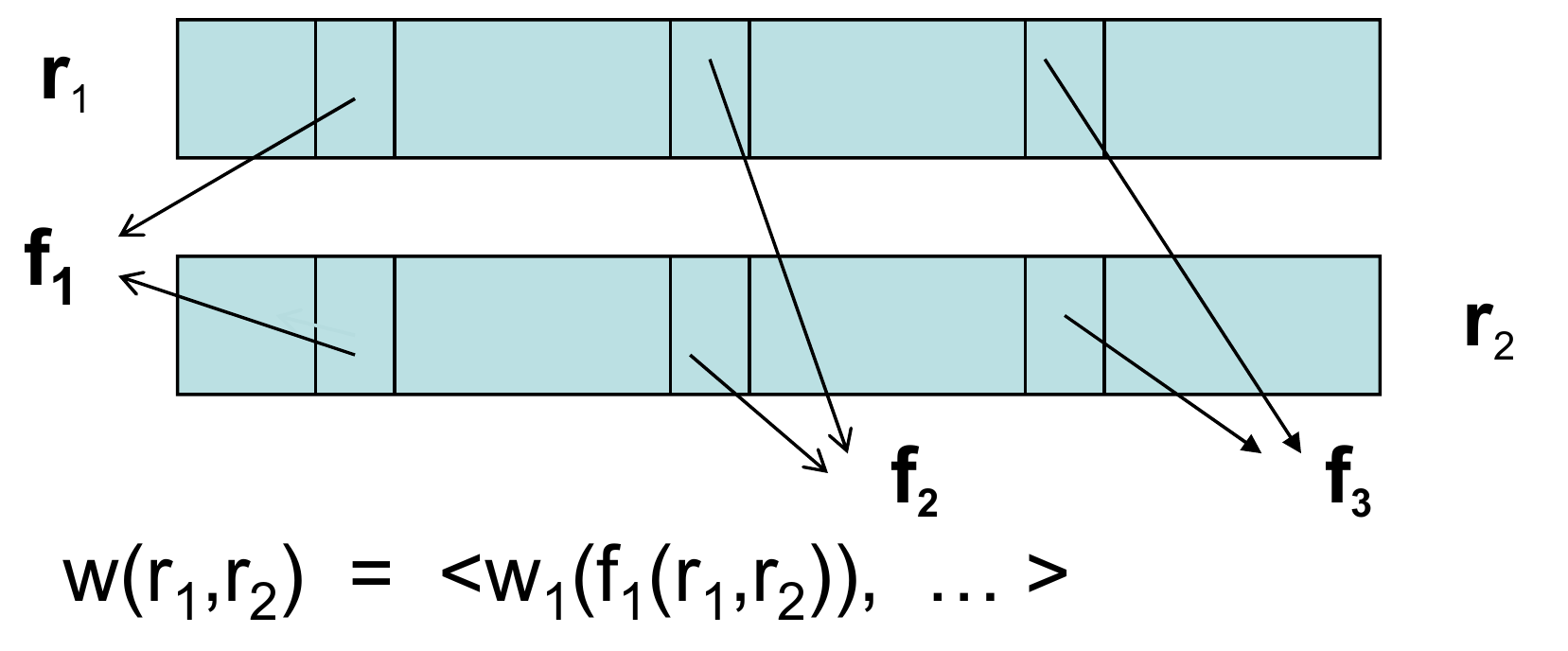}
\end{center}
\vspace{-4mm}
\caption{Feature-based similarity}\label{fig:sim}
 \vspace*{-5cm}
\end{center}
\end{wrapfigure}}

\begin{figure}[h]
\begin{center}
\vspace{3mm}
\begin{center}
\includegraphics[width=7cm]{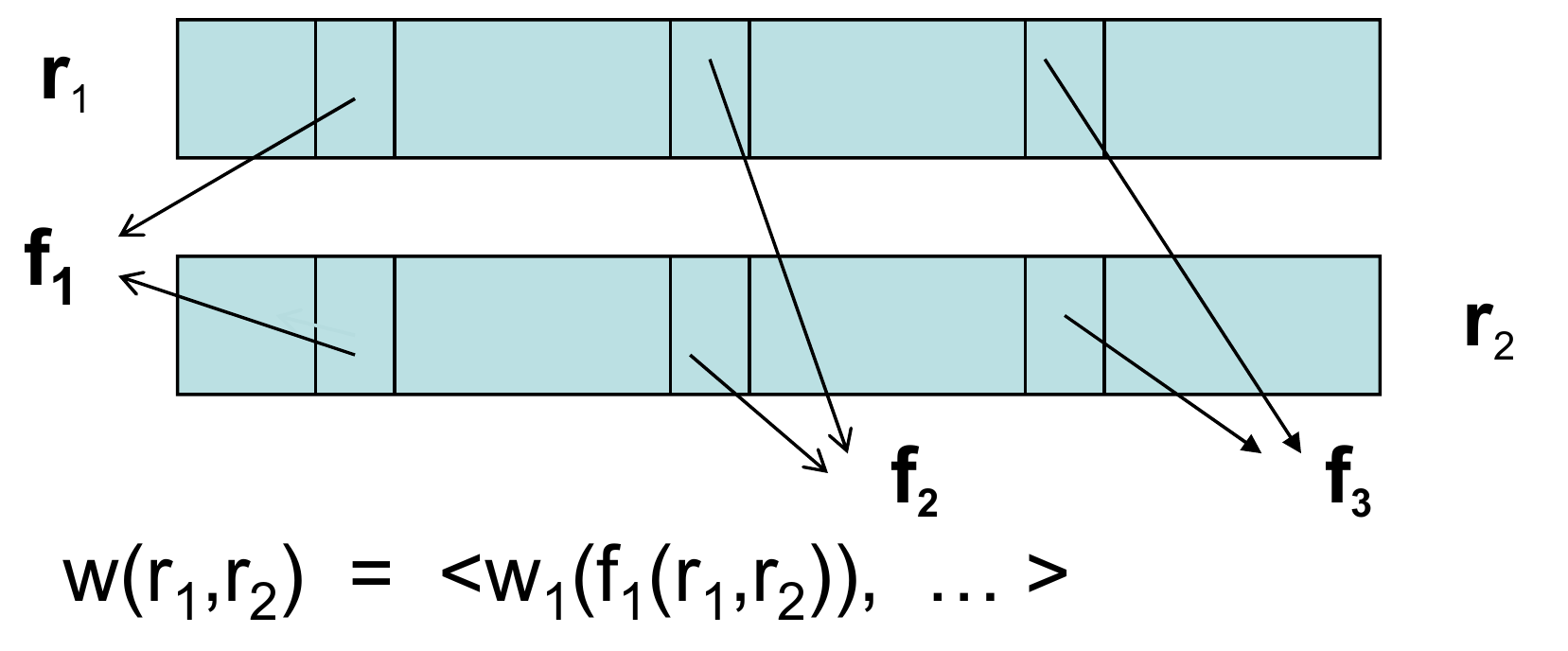}
\end{center}
\vspace{-4mm}
\caption{\ \ Feature-based similarity}\label{fig:sim}
\vspace{-6mm}
\end{center}
\end{figure}

Some record-pairs in the training dataset $T$ are considered as duplicates and others as non-duplicates, which results (according to path $\mf{4,7}$) in a  ``similarity-enhanced" training database $T^{\!s}$ of tuples of
the form $\langle r_1, r_2, w(r_1,r_2), L\rangle$, with label $L \in \{0,1\}$. Label $L$ indicates if the two records are duplicates ($L = 1$) or not ($L=0$). These labels are consistent with the corresponding weight vectors. The classifier is trained using $T^{\!s}$, leading, through the application of the SVM algorithm, to the classification model (edges $\mf{9,10}$) to be used for ER.

Blocking is applied to instance $D$, pre-classifying records into blocks, so that only records in a same block will form input pairs for the trained classification model. Accordingly,
two records in a same block may end up as duplicates (of each other) or not, but two records in different blocks will never be duplicates.

We assume each record $r \in D$ has two extra, auxiliary attributes: a unique and global (numerical) {\em record identifier} ({\em rid}) whose value is originally assigned and never changes;
and a {\em block number} that initially takes the {\em rid} as value. This block number is subject to changes.

For the records in $D$, similarity measures are used for blocking (see sub-path $\mf{5,8}$). To decide if two records, $r_1,r_2$, go into the same block, the weight-vector $w(r_1,r_2)$ can be used: it can be read off
from it if their values for certain attributes are similar enough or not. However, the similarity computations required for blocking
may be different from those involved in the computation of the weight-vectors $w(r_1,r_2)$, which are related to the classification model. Either way, this similarity information is used by  the {\em blocking-matching dependencies}, which are pre-declared and  domain-dependent.

Blocking-MDs specify and enforce (through their RHSs) that the blocks (block numbers) of two records have to be made identical.
This happens when certain similarities between pairs of attribute values appearing in the LHSs of the MDs hold. For example, (\ref{eq:block}) is a blocking-MD that requires the computation
 of similarities of string values for attribute $\nit{Title}$. The similarity-atoms on the LHS of a blocking-MD are considered to be true when the similarity values are above  thresholds
 that have been  predefined for blocking purposes only.

\ignore{
\begin{figure}[h]
\begin{center}
\vspace{-3mm}
\begin{center}
\includegraphics[width=7.5cm]{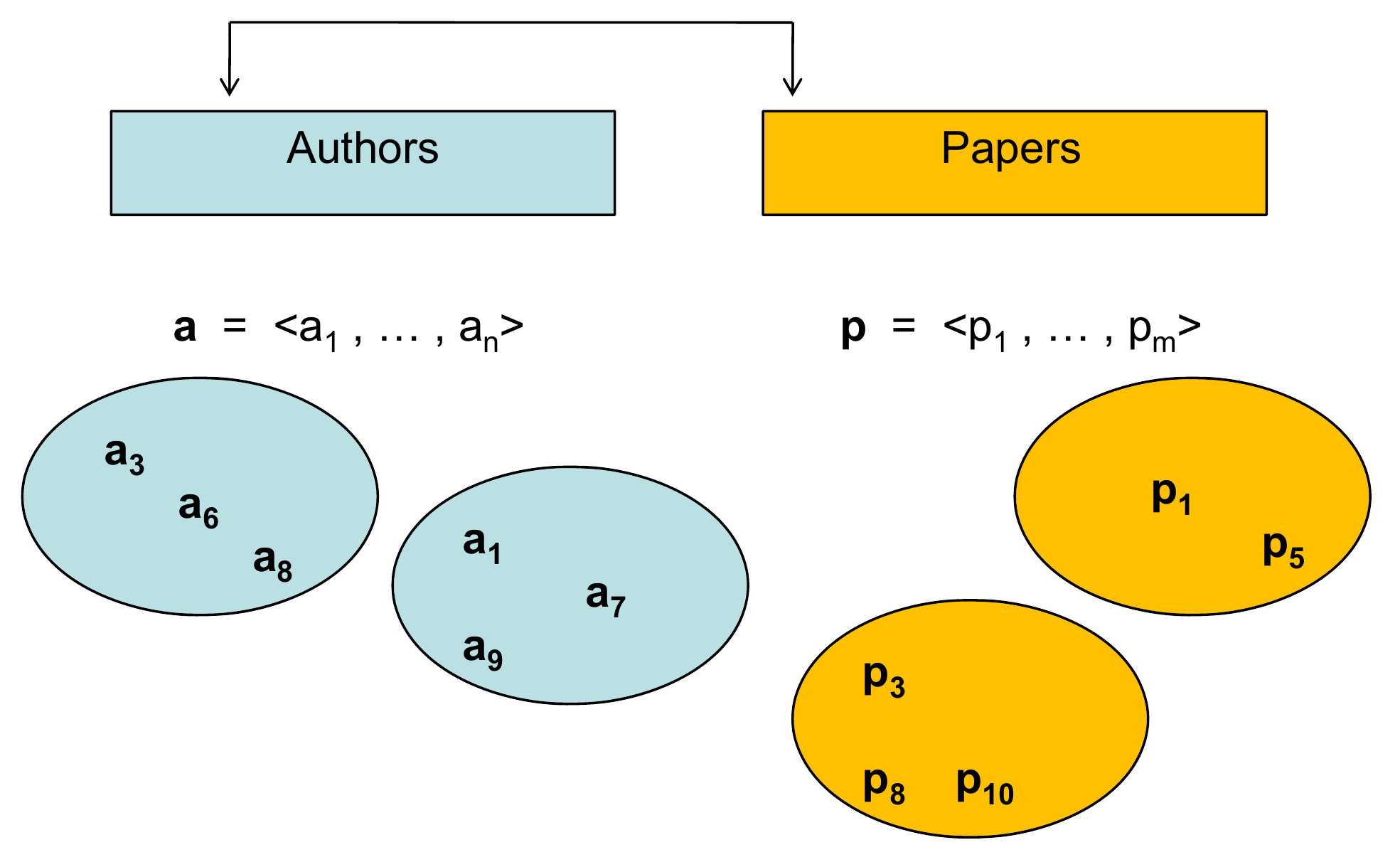}
\end{center}
\vspace{-4mm}
\caption{\ \ Collective blocking}\label{fig:blocks}
\vspace{-6mm}
\end{center}
\end{figure}
}

\begin{example} \label{ex:blockMD} (ex. \ref{ex:IntroBlock} cont.) \bl{With schema $\{\nit{Author(AID,Name,}$ $\nit{PTitle,}$ $\nit{ABlock)},$  $\nit{Paper(PID,PTitle,Venue,PBlock)}$ (including ID and block attributes), the following is a relational MD that captures a blocking policy that is similar to (but more refined than) that in Example \ref{ex:IntroBlock}:
{\small \begin{eqnarray}
\varphi\!: \ \ \nit{Author}(t_1,x_1,y_1,\nit{bl}_1) \  \wedge \ \nit{Paper}(t_3,y_1',z_1, \nit{bl}_4) \ \wedge \ y_1 \approx y_1' \ \wedge \hspace{4mm} \nonumber&&\\ \nit{Author}(t_2,x_2,y_2,\nit{bl}_2)  \ \wedge \  \nit{Paper}(t_4,y_2',z_2, \nit{bl}_4) \ \wedge \ y_2 \approx y_2' \ \wedge
 \nonumber  &&\\x_1 \approx x_2 \ \wedge \ y_1 \approx y_2    \longrightarrow \ \ \nit{bl}_1 \doteq \nit{bl}_2,~~~~~~ \label{eq:blmd}&&
 \end{eqnarray}}
\hspace*{-2mm} with  \nit{Author}-atoms as ``leading atoms" (they contain the identified variables on the RHS). It  contains   similarity comparisons involving  attribute values for both relations  \nit{Author} and \nit{Paper}.
It specifies that when the \nit{Author}-tuple similarities on the LHS hold, and their papers are similar to those in corresponding \nit{Paper}-tuples that are in the {\em same} block (equality as an implicit similarity is captured by the join variable $\nit{bl}_4$), then blocks $\nit{bl}_1, \nit{bl}_2$ have to be made identical. This blocking policy
uses relational knowledge (the relationships between \nit{Author} and \nit{Paper} tuples), plus the blocking decisions already made about \nit{Paper} tuples.}
\boxtheorem
\end{example}

We can see from (\ref{eq:blmd}) that information about  classifications in blocks of records for the entity at hand (\ent{Author} in this case) and for others entities (\ent{Paper} in this case) may
simultaneously appear as
conditions in the LHSs of blocking-MDs. Furthermore, blocking-MDs may involve in their LHSs inter-entity similarity conditions, such as $y_1 \approx y_1'$ in \re{(\ref{eq:blmd})}). \ignore{   similarity conditions about attribute values in records for entities that are different ($z_1 \approx_2 z_2$ for \ent{Paper} in \re{(\ref{eq:blmd})}) from that under blocking
(i.e.~\ent{Author}).} All this is the basis for our ``semantically-enhanced" collective blocking process.

The {\em MD-based collective blocking} stage (steps $\mf{5,8,6}$) consists in the enforcement of the blocking-MDs on $D$, which results in database $D$ enhanced with information about the blocks to which the records are assigned. Pairs of records with the same block form  {\em candidate duplicate record-pairs}.

\bl{We emphasize  that some blocking-MDs, such as (\ref{eq:blmd}), are more general than those of the form (\ref{eq:md2145})
 introduced in \cite{FanJLM09} or Section \ref{sec:mds}: In their LHSs, they may contain  regular database atoms, from more that one relation, that are used to give context to the similarity atoms
 in the MD, to capture additional relational knowledge. MDs of this kind are called {\em relational MDs}, and extend the so-called {\em classical MDs} of Section \ref{sec:mds}. (Cf. \ref{sec:relMDs} for more details
 on relational MDs.)}

\bl{A unique assignment of blocks to records is obtained after the enforcement of the blocking-MDs. Uniqueness is guaranteed by the properties of the class of MDs we use for
blocking. Actually, blocking-MDs will turn out to have the UCI property (cf.~Section \ref{sec:mds}). (More details on this are given in Sections \ref{MDBlocking} and \ref{sec:relMDs}.)}

After the records have been assigned to blocks, record-pairs $\langle r_1, r_2\rangle$, with $r_1,r_2$ in the same block, are considered for the duplicate test. At this point, we proceed as we did
for the training data: the weight-vectors $w(r_1,r_2)$, which represent the record-pairs in the ``feature vector space", are  computed and passed over to the classifier  (edges $\mf{11,12}$).\footnote{Similarity computations are kept in appropriate program predicates. So, similarity values computed before blocking can be reused at this stage, or whenever needed.} \ignore{ Next, tuples $\langle r_1, r_2,w(r_1,r_2)\rangle$ are used as input for the trained classification algorithm (edge $\mf{12}$).}

The result of  applying the  trained ML-based classifier to the record-pairs is a set of triples $\langle r_1,r_2,1\rangle$ containing records that come from the same block and are
considered to be duplicates. Equivalently, the output is a set $M \subseteq D\times D$ containing pairs of duplicate records (edge $\mf{13}$).\ignore{\footnote{The classifier also returns pairs or records that come from the same block, but are not considered to be duplicate. The set thereof in not
 interesting, at least as a workflow component.}} The records in pairs in $M$ are merged by enforcing an application-dependent set of (merge-)MDs (edge $\mf{14}$). This set of MDs is different from that used
 for blocking.

Since records have kept their {\em rids}, we define a ``similarity" predicate $\approx_\nit{id}$ on the domain of {\em rids} as follows: \ \
$r_1[\nit{rid}] \approx_\nit{id} r_2[\nit{rid}] \ \mbox{ iff } \
\langle r_1, r_2\rangle \in M$,
 i.e.~iff the corresponding records are considered to be duplicates by the classifier. We informally denote $r_1[\nit{rid}] \approx_\nit{id} r_2[\nit{rid}]$ by $r_1 \approx r_2$.
 Using this notation, the merge-MDs are usually and informally written in the form: \ \
$r_1 \approx r_2 \ \rightarrow \ r_1 \doteq r_2$. Here, the RHS is a shorthand for \ $r_1[A_1] \doteq r_2[A_1] \wedge \cdots \wedge r_1[A_m] \doteq r_2[A_m]$, where $A_1, \ldots, A_m$ are all the record attributes,
excluding the first and last, i.e.~ignoring the identifier and the block number (cf.~Figure~\ref{fig:recs}).
Putting all together, merge-MDs take the official form:
\begin{equation}
r_1[\nit{rid}] \approx_\nit{id} r_2[\nit{rid}] \ \longrightarrow \ r_1[A_1] \doteq r_2[A_1] \wedge \cdots \wedge r_1[A_m] \doteq r_2[A_m]. \label{eq:offBmds}
\end{equation}
Merging  at
the attribute level, as required by the RHS, uses the  predefined and domain-dependent matching functions $\match_{A_i}$.

After applying the merge-MDs, a single duplicate-free instance is obtained from $D$ (edge $\mf{15}$). Uniqueness is guaranteed by the fact that the classes of merge-MDs that we use in our
generic approach turn out to be interaction-free. (More details are given in Section \ref{Detection} and \ref{sec:relMDs}. See also the brief discussion in Section \ref{sec:mds}.)

More details about the {\em ERBlox} system and our approach to ER are found in the subsequent sections.

\section{Datasets and Similarity Computation}\label{sec:init}

We  now describe some aspects of the MAS dataset that are relevant for the description of the {\em ERBlox} system components,\footnote{We also independently experimented
with the DBLP and Cora Citation datasets, but we will concentrate on MAS.} and the way the initial data is processed and created for their use with the {\em LogiQL} language of {\em LogicBlox}.

\subsection{Data files  and relational data}

In the initial, structured data files, entries (non-relational records) for entity \ent{Author} relation contain author names and their affiliations. The entries for entity \ent{Paper} contain: paper titles, years of publication, conference IDs, journal IDs, and keywords. \bl{Entries for the \ent{PaperAuthor} relationship between \ent{Paper} and \ent{Author} entities} contain: paper IDs, author IDs, author names, and their affiliations. The entries for the \ent{Journal} and \ent{Conference} entities contain both short names of the publication venue, their full names, and their home pages.

The dataset is preprocessed by means of Python scripts, in preparation for proper {\em ERBlox} tasks. This is necessary because the data gathering methods in general, and for the MAS dataset in particular, are often loosely controlled, resulting in out-of-range values, impossible data combinations, missing values, etc. For example, non-word characters are replaced by blanks, some strings are converted into lower case, etc. Not solving
these problems may lead to later execution problems and, in the end, to   misleading ER results. This preprocessing produces updated structured data files. As expected, there is no ER at this stage, and in the new files
there may be many authors who publish under several variations of their names; also the same paper may appear under slightly different titles, etc. This kind of cleaning  will be performed with {\em ERBlox}.


\ignore{
\red{Data in the dataset is pre-processed by Python scripts before performing ER.
This is because data gathering methods are often loosely controlled, resulting in out-of-range values, impossible data combinations, missing values, etc. Analyzing data that has not been carefully screened for such problems can produce misleading ER results. For instance, we pre-process data by replacing all non-word characters with blanks and
converting strings to lowercase.}
\comlb{The paragraph in blue below is misleading. It may sound as if your are already doing ER at this stage, as part of {\em ERBlox}, which is not as far as I can see ... It has to be made more clear.}
\blue{
By using {\em ERBlox} on this dataset, we determine which papers in MAS data are written by a given author. This is clear case of ER since there are many authors who publish under several variations of their names. Also the same paper may appear under slightly different titles, etc. }
}


Next, from the data \ignore{dataset, which contains the data} in (the preprocessed) structured files, relational predicates and their extensions are created and computed,
 by means \ignore{for intentional, relational predicates are computed by} of a generic Datalog program in {\em LogiQL}  \cite{Aref15,Halpin15}. For example, these rules are part of the program:
\begin{eqnarray}
&&\hspace*{-0.7cm}\_\nit{file}\_\nit{in}(x1,x2,x3) \ \longrightarrow \ \nit{string}(x1),\nit{string}(x2), \nit{string}(x3). \label{eq:one}\\
&&\hspace*{-0.6cm}\nit{lang:physical:filePath}[`\_\nit{file}\_\nit{in}] = "\nit{author.csv}". \label{eq:two}\\
&&\hspace*{-1.3cm}\nit{+author}(\nit{id}1,x2,x3) \ \leftarrow  \_\nit{file}\_\nit{in}(x1, x2,x3),
\nit{string\!:\!int}64\!:\!\nit{convert}[x1] = \nit{id}1. \label{eq:three}
\end{eqnarray}
Here, (\ref{eq:one}) is a predicate schema declaration, in this case of  the ``$\_$\ent{file}$\_$\ent{in}" predicate with three string-valued attributes. It is used to automatically store the contents extracted from the source file "author.csv", as specified in (\ref{eq:two}). In {\em LogiQL} in general, metadata declarations use ``$\rightarrow$".
(In {\em LogiQL}, each predicate's schema has to be declared,  unless it can be inferred from the rest of the program.)
 Derivation rules, such as (\ref{eq:three}), use ``$\leftarrow$", as usual in Datalog. It defines the \ent{author} predicate, and the ``$+$" in the rule head inserts the data into the predicate extension.
The rule also makes the first attribute a tuple identifier.

Figure~\ref{fig:dataset} shows three relational predicates that are created and populated in this way: $\nit{Author(AID, Name,Affiliation, Bl\#})$, $\nit{Paper(PID,Title,Year,CID},$ $\nit{JID,Keyword,Bl \#})$,
$\nit{PaperAuthor(PID,AID,Name,Affiliation})$. The (partial) tables show that there may be missing attribute values.

\begin{figure}[t]
{\scriptsize
\begin{center}\begin{tabular}{r|r|l|l|r|}\hline
$\nit{Author}$& $\nit{AID}$ &$\nit{Name}$ & $\nit{Affiliation}$ & $Bl \#$  \\ \hline
$$ &$659$&$\nit{Jean}\mymathhyphen{Pierre\ Olivier\ de}$ & $\nit{Ecole\ des\ Hautes}$ & $659$  \\
$$ &$2546$&$\nit{Olivier\ de\ Sardan}$ & $\nit{Recherche\ Scientifique}$ & $2546$  \\
$$ &$612$&$\nit{Matthias\ Roeckl}$ & $\nit{German\ Aerospace\ Center}$ & $612$ \\
$$ &$4994$&$\nit{Matthias\ Roeckl}$ & $\nit{Institute\ of\ Communications}$ & $4994$ \\\cline{2-5}
\end{tabular}\end{center}
\ignore{\begin{center}\begin{tabular}{r|r|l|c|c|c|l|c|}\hline
$\nit{Paper}$&  $\nit{PID}$ &$\nit{Title}$ & $\nit{Year}$  & $CID$ & $JID$&$Keyword$& $Bl \#$\\ \hline
$$ &$123$&$\nit{Illness\ entities\ in\ West\ Africa}$ & $1998$ & $179$& $$ & $\nit{West\ Africa,\ Illness}$ & $123$\\
$$ &$205$&$\nit{Illness\ entities\ in\ Africa}$ & $1998$  & $179$ &  $$ &$\nit{Africa, Illness}$& $205$\\
$$ &$769$&$\nit{DLR\ Simulation\ Environment\ m3}$ & $2007$  & $146$ &  $$ &$\nit{ Simulation\ m3}$& $769$\\
$$ &$195$&$\nit{DLR\ Simulation\ Environment}$ & $2007$  & $146$ &  $$ &$\nit{ Simulation}$& $195$\\\cline{2-8}
\end{tabular}\end{center}}

\begin{center}\begin{tabular}{r|r|l|c|c|c|c} \hhline{------~}
$\nit{Paper}$&  $\nit{PID}$ &$\nit{Title}$ & $\nit{Year}$  & $CID$ & $JID$&$\cdots$\\ \hhline{------~}
$$ &$123$&$\nit{Illness\ entities\ in\ West\ Africa}$ & $1998$ & $179$& $$ & \\
$$ &$205$&$\nit{Illness\ entities\ in\ Africa}$ & $1998$  & $179$ &  $$ &\\
$$ &$769$&$\nit{DLR\ Simulation\ Environment\ m3}$ & $2007$  & $146$ &  $$ &\\
$$ &$195$&$\nit{DLR\ Simulation\ Environment}$ & $2007$  & $146$ &  $$ &$\cdots$\\\cline{2-6}
\end{tabular}\end{center}
\vspace{-3mm}\hspace*{6.7cm}\begin{tabular}{c|l|c|}\hhline{~--}
 $\cdots$ &$Keyword$& $Bl \#$\\ \hhline{~--}
 & $\nit{West\ Africa,\ Illness}$ & $123$\\
 &$\nit{Africa, Illness}$& $205$\\
&$\nit{ Simulation\ m3}$& $769$\\
$\cdots$&$\nit{ Simulation}$& $195$\\\cline{2-3}
\end{tabular}

\begin{center}\begin{tabular}{c|c|r|l|l|}\hline
$\nit{PaperAuthor}$&$\nit{PID}$ & $\nit{AID}$ & $\nit{Name}$ & $\nit{Affiliation}$ \\ \hline
$$&$123$ & $659$ & $\nit{Jean}\mymathhyphen\nit{Pierre\ Olivier\ de}$ & $\nit{Ecole\ des\ Hautes}$  \\
$$&$205$ & $2546$ & $\nit{Olivier\ de\ Sardan}$ & $\nit{Recherche\ Scientifique}$ \\
$$&$769$ & $612$ & $\nit{Matthias\ Roeckl}$ & $\nit{German\ Aerospace\ Center}$ \\
$$&$195$ & $4994$ & $\nit{Matthias\ Roeckl}$ & $\nit{Institute\ of\ Communications}$ \\ \cline{2-5}
\end{tabular}\end{center}
\ignore{\begin{center}\begin{tabular}{r|r|l|l|l|r|}\hline
$\nit{Journal}$& $\nit{CID}$ &$\nit{SName}$ & $\nit{FName}$ & $\nit{HPage}$ & $\nit{Bl\#}$ \\ \hline
$$ &$189$&$\nit{F.\ Cass}$ & $\nit{Frank\ Cass}$ & $$ & $189$  \\
$$ &$152$&$\nit{INTERNET}$ & $\nit{IEEE\ Internet\ Computing}$ & $\nit{computer.org/internet}$ & $152$ \\ \cline{2-6}
\end{tabular}\end{center}
\begin{center}\begin{tabular}{r|r|c|l|c|c|}\hline
$\nit{Conference}$& $\nit{CID}$ &$\nit{SName}$ & $\nit{FName}$ & $\nit{HPage}$ & $\nit{Bl\#}$  \\ \hline
 $$ &$179$&$$ & $\nit{Medical\ Anthropology }$ & $\nit{medant.com}$ & $179$ \\
 $$ &$146$&$$ & $\nit{First\ C2C}\mymathhyphen\nit{CC}\mymathhyphen\nit{COMeSafety\ Simulation}$ & $$ & $146$ \\ \cline{2-6}
\end{tabular}\end{center}
\begin{center}\begin{tabular}{c|l|l|}\hline
$\nit{CoAuthor}$&$\nit{AID}_1$ & $\nit{AID}_2$  \\ \hline
$$&$5026$ & $659$  \\
$$&$5026$ & $2546$ \\ \cline{2-3}
\end{tabular}\end{center}}}
\vspace{-4mm}\caption{Relation extensions from MAS using  LogiQL rules} \label{fig:dataset} \vspace{-4mm}
\end{figure}

\subsection{Features and similarity computation}\label{sec:features}

From the general description of our methodology in Section \ref{MLMDFramework}, a crucial component is {\em similarity computation}. It is needed for: \ (a) blocking, and \ (b) building the classification model.
Similarity measures are related to {\em features}, which are numerical functions of the data,  more precisely of the values of some specially chosen attributes. Feature selection is a fundamental
task in machine learning \cite{dash,tang}; going in detail into this subject is beyond the scope of this work. \ Example \ref{ex:similarityComputation} shows some specific aspects of this task as related to our dataset.


In relation to blocking, in order to decide if two records, $r_1,r_2$ in $D$, go into the same block, similarity of values for certain attributes are computed,  those that are appear in similarity
conditions in the LHSs of blocking-MDs.
All is needed is whether they are similar enough or not, which is determined by predefined numerical thresholds.

For model building, similarity values are computed to build the weight-vectors, $w(r_1,r_2)$, for records $r_1, r_2$ from the training data in $T$.
The numerical values in those vectors depend on the values taken by some selected record attributes  (cf.~Figure~\ref{fig:sim}).

\begin{example}\label{ex:similarityComputation} (ex. \ref{ex:IntroBlock} cont.) Bibliographic datasets, such as MAS, have been  commonly used for evaluation of  machine learning techniques, in particular, classification for ER. In our case, the features chosen in our work for the classification of records for entities \ent{Paper} and \ent{Author} from the MAS dataset (and the other datasets)  correspond to
those previously used in \cite{Torvik09, Christen2008}. Experiments in \cite{Kopcke08} show that the chosen features enhance generalization power of the classification model, by reducing over-fitting.

In the case of \ent{Paper}-records, if the ``journal ID" values are null in both records, but not their ``conference ID" values,  ``journal ID" is not considered
for feature computation, because it does not contribute to the recall or precision of the classifier under construction.  Similarly, when the ``conference ID" values are null. However, the values for  ``journal ID" and ``conference ID" are replaced by  ``journal full name" and ``conference full name" values that are
 found in \ent{Conference}- and \ent{Journal}-records, resp. Attributes \nit{Title}, \nit{Year}, \nit{ConfFullName} or \nit{JourFullName}, and \nit{Keyword} are chosen for feature computation.

For feature computation in the case of  \ent{Author}-records, the \nit{Name} attribute is split in two, the \nit{Fname} and \nit{Lname} attributes, to increase recall and precision of the classifier under construction. Accordingly, features are computed
for attributes \nit{Fname}, \nit{Lname} and \nit{Affiliation}. \boxtheorem
\end{example}

Once the classifier has been built, also weight-vectors, $w(r_1,r_2)$ are computed as inputs
for the classifier, but this time
for records from the data under classification (in $D$).\footnote{In our experiments, we did not care about null values in records under classification. Learning, inference, and prediction in the presence of missing values are pervasive problems in machine learning and statistical data analysis.  Dealing with missing values is beyond the scope of this work.}

Notice that numerical values, associated to similarities, in a weight-vector $w(r_1,r_2)$ for $r_1, r_2$ under classification, could be used as similarity information for blocking.
However, the attributes and features used for blocking
may be different from those used for weight-vectors. For example, in our experiments with the MAS dataset, the classification of \ent{Author}-records is based on attributes \nit{Fname}, \nit{Lname}, and \nit{Affiliation}.
For blocking, the latter is reused as such (cf.~MD (\ref{eqq:md2}) below), but also the combination of \nit{Fname} and \nit{Lname} is reused, as attribute \nit{Name} in MDs (cf.~MDs (\ref{eqq:md2}) and (\ref{eqq:md4}) below).


There is a class of well-known and widely applied similarity functions that are used in data cleaning and machine learning \cite{cohen03}. For our application with {\em ERBlox} we
 used three of them, depending on the attribute domains for the MAS dataset. Long-text-valued attributes, in our case, e.g. for the {\em Affiliation} attribute, their values are represented as lists of strings.
  For computing similarities between these kinds of attribute values, the ``TF-IDF cosine" measure was used \cite{Salton88}.  It assigns low weights to frequent strings and high weights to rare strings. For example,  affiliation values usually contain multiple strings, e.g. ``Carleton University, School of Computer Science". Among them, some are frequent, e.g. ``School", and others are rare, e.g. ``Carleton".

For attributes with ``short" string values, such as author names,  ``Jaro-Winkler" similarity was used \cite{Jaro95,Winkler99}. This measure counts the characters in common in two strings, even if they are misplaced by a short distance. For example, this measure gives a high similarity value to the pair of first names ``Zeinab" and ``Zienab". In the MAS dataset, there are many author first names and last names presenting this kind of
misspellings.

For numerical attributes, such as publication year, the ``Levenshtein distance" was used \cite{Navarro}. The similarity of two numbers is based on the minimum number of operations required to transform one into the other.

As already mentioned in Section \ref{MLMDFramework},
these similarity measures are used, but differently, for blocking and  the creation and application of the classification algorithm. In the former case, similarity values related to LHSs of blocking-MDs are
compared with user-defined thresholds, in essence, making them boolean variables. In the latter case, they are used for computing the similarity vectors, which contain numerical values (in $\mathbb{R}$). Notice
 that similarity measures are {\em not} used beyond the output of the classification algorithm, in particular, not for
MD-based record merging.

Similarity computation for {\em ERBlox}
is done through {\em LogiQL}-rules that define the similarity functions. In particular, similarity computations are kept in extensions of program-defined predicates. For example, if the similarity value for the pair of
values, $a_1,a_2$, for
attribute $\nit{Title}$ is above the threshold, a tuple $\nit{Title\mbox{{\it -}}Sim}(a_1,a_2)$ is created by  the program. 

\section{MD-Based Collective Blocking} \label{MDBlocking}

\bl{As described in Section \ref{MLMDFramework}, the \nit{Block} attribute, $\nit{Bl}$, in records takes integer numerical values; and before the blocking process starts (or blocking-MDs are enforced), each record in the instance $D$ has a unique block number that coincides with its {\em rid}. Blocking policies are specified by blocking-MDs, all of which use the same matching function for identity enforcement, given by:}
\red{\begin{equation}
\mbox{For } i, j \in \mathbb{N}, \mbox{ with } j \leq i, \ \ \match_\nit{Bl}(i, j):= i. \label{eq:mf}
\end{equation}}
A blocking MD that identifies block numbers (i.e.~makes them identical) in two records (tuples) for database relation $R$ (cf.~Figure~\ref{fig:recs}) takes the form:
\begin{equation}
R(\bar{x}_1,\nit{bl}_1) \ \wedge \ R(\bar{x}_2,\nit{bl}_2) \ \wedge \ \psi(\bar{x}_3) \ \longrightarrow \ \nit{bl}_1\doteq \nit{bl}_2. \label{eq:newMDs}
\end{equation}
\bl{Here, $\nit{bl}_1, \nit{bl}_2$ are variables for block numbers, $R$ is a database predicate (representing an entity), the lists of variables $\bar{x}_1, \bar{x}_2$ stand for all the attributes in $R$ but $\nit{Bl\#}$,
for which variables $\nit{bl}_i$ are used. The MD in (\ref{eq:newMDs}) is {\em relational} when formula $\psi$ in it is a conjunction of relational atoms plus comparison atoms via similarity predicates;
  including implicit equalities of block numbers (but not $\approx$-similarities between block numbers).  The variables in $\psi(\bar{x}_3)$ may appear among those in $\bar{x}_1, \bar{x}_2$ (in $R$) or in another database predicate or in a similarity atom. We assume that $(\bar{x}_1 \cup \bar{x}_2) \cap \bar{x}_3 \neq\emptyset$. \ \bl{(Cf. \ref{sec:relMDs} for more details on relational MDs.)}}

 \bl{   An example is the MD in (\ref{eq:blmd}), where the leading $R_1, R_2$-atoms are \ent{Author} tuples, the extra conjunction contains \ent{Paper} atoms, non-block-similarities, and an implicit equality of blocks through the shared use of variable $\nit{bl}_4$. There,  $\psi$ is $\nit{Paper}(t_3,y_1',z_1, \nit{bl}_4) \ \wedge \ y_1 \approx y_1' \ \wedge \  \nit{Paper}(t_4,y_2',z_2, \nit{bl}_4) \ \wedge \ y_2 \approx y_2' \ \wedge
\ x_1 \approx x_2 \ \wedge \ y_1 \approx y_2$. }

\begin{example}\label{ex:blockLoqic} These are some of the blocking-MDs used with the MAS dataset. \bl{The first two are {\em classical} blocking-MDs, and the last two are properly
{\em relational} blocking-MDs:}
\begin{eqnarray}
&&\hspace*{-0.8cm}  \nit{Paper}(\nit{pid}_1,x_1,y_1,z_1,w_1,v_1,\nit{bl}_1) \ \wedge \ \nit{Paper}(\nit{pid}_2,x_2,y_2,z_2,w_2,v_2,\nit{bl}_2) \ \wedge \label{eqq:md1}\\
&& \hspace*{3cm} x_1 \approx_{\!\nit{Title}} x_2 \ \wedge \ y_1=y_2 \ \wedge z_1=z_2 \ \longrightarrow \  \ \nit{bl}_1 \doteq \nit{bl}_2. \nonumber\\
&&\hspace*{-0.8cm}  \nit{Author}(\nit{aid}_1, x_1,y_1, \nit{bl}_1) \ \wedge \ \nit{Author}(\nit{aid}_2, x_2,y_2, \nit{bl}_2) \ \wedge \label{eqq:md2} \\
&&\hspace*{4cm}x_1 \approx_{\!\nit{Name}} x_2 \ \wedge \ y_1\approx_{\!\nit{Aff}}y_2 \ \longrightarrow \  \ \nit{bl}_1 \doteq \nit{bl}_2.\nonumber
\end{eqnarray}\vspace{-12mm}
\begin{eqnarray}
&&\hspace*{-0.8cm}  \nit{Paper}(\nit{pid}_1,x_1,y_1,z_1,w_1,v_1,\nit{bl}_1) \ \wedge \ \nit{Paper}(\nit{pid}_2,x_2,y_2,z_2,w_2,v_2,\nit{bl}_2) \wedge \label{eqq:md3} \\
&&\hspace*{0.05cm}\nit{PaperAuthor}(\nit{pid}_1, \nit{aid}_1, x'_1,y'_1) \ \wedge \nit{PaperAuthor}(\nit{pid}_2, \nit{aid}_2, x'_2,y'_2) \ \wedge \nonumber \\
&&\hspace*{0.05cm} \nit{Author}(\nit{aid}_1, x'_1,y'_1, \nit{bl}_3) \ \wedge \nit{Author}(\nit{aid}_2, x'_2,y'_2, \nit{bl}_3) \wedge x_1 \approx_{\!\nit{Title}} x_2 \nonumber \\   &&\hspace*{8.3cm} \longrightarrow \ \nit{bl}_1 \doteq \nit{bl}_2.\nonumber
\end{eqnarray}\vspace{-12mm}
\begin{eqnarray}
&&\hspace*{-0.8cm} \nit{Author}(\nit{aid}_1, x_1,y_1, \nit{bl}_1) \ \wedge \nit{Author}(\nit{aid}_2, x_2,y_2, \nit{bl}_2) \ \wedge \ x_1 \approx_{\!\nit{Name}} x_2\wedge  \label{eqq:md4}\\
&&\hspace*{0.05cm}
 \nit{PaperAuthor}(\nit{pid}_1, \nit{aid}_1, x_1,y_1) \ \wedge \nit{PaperAuthor}(\nit{pid}_2, \nit{aid}_2, x_2,y_2) \wedge \nonumber\\
 &&\hspace*{0.05cm}\nit{Paper}(\nit{pid}_1,x'_1,y'_1,z'_1,w'_1,v'_1,\nit{bl}_3)  \wedge  \nit{Paper}(\nit{pid}_2,x'_2,y'_2,z'_2,w'_2,v'_2,\nit{bl}_3)\nonumber\\&&\hspace*{8cm}  \ \longrightarrow \   \nit{bl}_1 \doteq \nit{bl}_2.\nonumber
\end{eqnarray}
In informal terms, (\ref{eqq:md1}) requires that, for every two \ent{Paper} entities $\mathbf{p}_1, \mathbf{p}_2$  for which the values
for attribute $\nit{Title}$ are similar, and with the same publication year and conference ID,  the values for attribute
${\it Bl\#}$ must be made identical. According to (\ref{eqq:md2}), whenever there are similar values for name and affiliation in \ent{Author}, the corresponding authors should go into the same block.

\bl{The relational blocking-MDs in  (\ref{eqq:md3}) and (\ref{eqq:md4})
 {\em collectively block} \ent{Paper} and \ent{Author} entities.} According to  (\ref{eqq:md3}), a blocking-MD for  \ent{Paper}, if two authors are in the same block, their papers $\mathbf{p}_1$, $\mathbf{p}_2$ having similar titles must be in the same block too. Notice that if papers $\mathbf{p}_1$ and $\mathbf{p}_2$ have similar titles, but they do not have same publication year or conference ID, we cannot block them together using (\ref{eqq:md1}) alone. The blocking-MD (\ref{eqq:md4}) for \ent{Author} is similar to that discussed in Example \ref{ex:blockMD}. \boxtheorem
\end{example}

For the application-dependent set, $\Sigma^\nit{Bl}$, of blocking-MDs we adopt the  chase-based semantics \cite{Bertossi12}, which may lead, in general, to several, alternative
final
instances. In each of them, every record is assigned to a unique block, but now records may share block numbers, which is interpreted as belonging to the same block. In principle, there might be two final instances where the same pair of records is put in the same block in one of them, but not in the other one.
\red{However, with a set of the relational blocking-MDs of the form (\ref{eq:newMDs}) acting on an initial instance $D$ (created with {\em LogicBlox} as described above), the chase-based enforcement of the MDs  results in a single, final instance, $D^{\!\nit{Bl}}$.
This is because the combination of the blocking-MDs with the initial instance $D$ turns out to belong to the SFAI class, which has the UCI property (cf.~Section \ref{sec:mds} and \ref{sec:relMDs}). }

\bl{That the initial instance and the blocking-MDs form a SFAI combination is easy to see. In fact, initially the block numbers in tuples (or records) are all different, they are the same as their tids.
 Now, the only relevant attributes in records (for SFAI membership) are ``block attributes", those appearing in RHSs of blocking-MDs (cf.~(\ref{eq:newMDs})). In the LHSs of blocking-MDs they may appear only in implicit equality atoms. Since all initial block numbers in $D$ are different,
no relevant similarity holds in $D$.}

\bl{Due to the SFAI property of  blocking-MDs in combination with the initial instance, MD enforcement leads to a single instance that  can be computed in polynomial time in data, which gives us the hope to
use a computationally well-behaved extension of plain Datalog for MD enforcement (and blocking).  It turns out that the representation and enforcement of these MDs can be done by means of Datalog with stratified negation
 \cite{ceri90,ahv95}, which is supported by {\em LogiQL}. Stratified Datalog programs have a  unique stable model, which
can be computed  in a bottom-up manner in polynomial time in the size of the extensional database.\footnote{General sets of MDs can be specified and enforced by means of disjunctive, stratified answer set programs, with \red{the possibly} multiple
 resolved instances corresponding to the stable models of the program  \cite{Bahmani12}. These programs can be {\em specialized}, via an automated rewriting mechanism, for the SFAI case, obtaining residual programs in Datalog with stratified negation
 \cite{BahmaniExten12}.}}

 In {\em LogiQL}, blocking-MDs take the form as Datalog rules:
\begin{eqnarray}
&&\hspace*{-0.8cm} R[\bar{X}_1]\!=\!\nit{Bl}_2 , \ \ R[\bar{X}_2]\!=\!\nit{Bl}_2  \ \ \longleftarrow \ \ \ R[\bar{X}_1]=\nit{Bl}_1, \ \ R[\bar{X}_2]=\nit{Bl}_2, \label{LQmd}\\ &&\hspace{5cm}  \psi(\bar X_3), \ \ \nit{Bl}_1 < \nit{Bl}_2, \nonumber
\end{eqnarray}
subject to the same conditions as for  (\ref{eq:newMDs}). \red{The condition $\nit{Bl}_1 < \nit{Bl}_2$ in the rule body corresponds to the use of the MF $\match_\nit{Bl}$ in (\ref{eq:mf})}.


An atom of the form $R[\bar{X}]\!\!=\!\!\nit{Bl}$ not only declares $\nit{Bl}$ as an attribute value for $R$, but also that predicate $R$ is functional on $\bar{X}$ \cite{Aref15}:  Each record in $R$ can have only one block number.

In addition to the blocking-MDs, we need some auxiliary rules, which we introduce and discuss next.
Given an initial instance $D$ and a set of blocking-MDs $\Sigma^\nit{Bl}$, the {\em LogiQL}-program $\Pi^{\!\nit{Bl}}(D)$ that specifies MD-based collective blocking contains the following rules:

\begin{itemize}
 \item[{\bf 1.}]
For every atom $R(\nit{rid}, \bar{x}, \nit{bl})$ $\in$ $D$, the fact  $R[\nit{rid}, \bar{x}]=\nit{bl}$. \ That is, initially, the block number, $\nit{bl}$, is functionally assigned  the value $\nit{rid}$.

\item[{\bf 2.}]
Facts of the form $\mbox{\nit{A-Sim}}(a_1,a_2)$, where
$a_1, a_2 \in \nit{Dom}(A)$, the finite attribute domain of an attribute $A$. They state that the two values are similar, which is determined by similarity computation. (Cf.~ Section
\ref{sec:features} for more on similarity computation.)

\ignore{\comlb{What is the connection between these similarity facts and the similarity measures discussed in the previous section, and the features discussed in the next section? I would guess
there is a common base and reuse.}  }

\item[{\bf 3.}] Rules for the blocking-MDs, as in  (\ref{LQmd}).

\item[{\bf 4.}]  Rules specifying older versions of entity records (in relation $R$) after MD-enforcement:
$$R\mbox{-}\nit{OldVer}(r,\bar{x}, \nit{bl}_1)  \ \ \longleftarrow \ \  R[r, \bar{x}]=\nit{bl}_1, \ R[r, \bar{x}]=\nit{bl}_2, \ \nit{bl}_1 < \nit{bl}_2.$$
Here, variable $r$ stands for the {\em rid}. Since for each {\it rid}, $r$, there could be several
atoms of the form  $R[r,\bar x]\!=\! \nit{bl}$,  corresponding to the evolution of the record identified by $r$ through an MD-based chase sequence, the rule specifies as old those versions of the record with a block number
that is smaller than the last one obtained for it.

\item[{\bf 5.}] Rules that collect the records' latest versions, to form blocks:
$$R\mbox{-}\nit{Block}[r,\bar{x}]= \nit{bl} \;\;\longleftarrow \;\;R[r,\bar{x}]= \nit{bl}, \ \nit{not} \ R\mbox{-}\nit{OldVer}(r,\bar{x}, \nit{bl}).$$
 The rule collects $R$-records that are not old versions.\footnote{{\em LogiQL}, uses  ``!" instead of $\nit{not}$ for Datalog negation \cite{Aref15}.}
\end{itemize}

Program $\Pi^{\!\nit{Bl}}(D)$ as above is a  Datalog program with stratified negation (there is no recursion through negation). In computational terms, this means that the program computes
old version of records (using negation), and next definitive blocks are computed. As expected from the SFAI property of blocking-MDs in combination with the initial instance, the program has and
computes a single model, in polynomial time in the size of the initial instance. From it, the final block numbers of records can be read off.

\begin{example}\label{LP} (ex. \ref{ex:blockLoqic} cont.) \ We consider only blocking-MDs (\ref{eqq:md1}) and (\ref{eqq:md3}).  The portion of $\Pi^{\!\nit{Bl}}(D)$ that does the blocking of records for the \ent{Paper} entity has the following rules (we follow the numbering used in the generic program):
\begin{itemize}
\item [{\bf 2.}] Facts such as: \\
$\nit{Title\mbox{{\it -}}Sim}(\nit{``Illness\ entities\ in\ West\ Africa"},\nit{``Illness\ entities\ in\ Africa"})$.\\ $\nit{Title\mbox{{\it -}}Sim}(\nit{``DLR\ Simulation\ Environment\ m3"}, \nit{``DLR\ Simulation}$ \\ 
\phantom{po}\hfill $\nit{Environment"}).$

\item [{\bf 3.}] $\nit{Paper}[\nit{pid}_1,x_1,y_1,z_1,w_1,v_1]=\nit{bl}_2, \nit{Paper}[\nit{pid}_2,x_2,y_2,z_2,w_2,v_2]=\nit{bl}_2 \ \leftarrow$

\vspace{-1mm}
\hfill $\nit{Paper}[\nit{pid}_1,x_1,y_1,z_1,w_1,v_1]=\nit{bl}_1, \nit{Paper}[\nit{pid}_2,x_2,y_2,z_2,w_2,v_2]=\nit{bl}_2,$

\vspace{-1mm}
\hfill$\nit{Title\mbox{{\it -}}Sim}(x_1,x_2), y_1=y_2, z_1=z_2, \nit{bl}_1 < \nit{bl}_2.$

\vspace{1mm}
$\nit{Paper}[\nit{pid}_1,x_1,y_1,z_1,w_1,v_1]=\nit{bl}_2, \nit{Paper}[\nit{pid}_2,x_2,y_2,z_2,w_2,v_2]=\nit{bl}_2\ \leftarrow$

\vspace{-1mm}
 \hspace*{0.4cm} $\nit{Paper}[\nit{pid}_1,x_1,y_1,z_1,w_1,v_1]=\nit{bl}_1, \nit{Paper}[\nit{pid}_2,x_2,y_2,z_2,w_2,v_2]=\nit{bl}_2,$

\vspace{-1mm}
 \hspace*{2.1cm}$  \nit{Title\mbox{{\it -}}Sim}(x_1,x_2), \nit{PaperAuthor}(\nit{pid}_1,\nit{aid}_1, x'_1,y'_1), \nit{bl}_1 < \nit{bl}_2, $

\vspace{-1mm}
 \hspace*{2.4cm}$\nit{PaperAuthor}(\nit{pid}_2,\nit{aid}_2, x'_2,y'_2), \nit{Author}[\nit{aid}_1, x'_1,y'_1]=\nit{bl}_3,$

\vspace{-1mm}
\hspace*{7.1cm} $\nit{Author}[\nit{aid}_2, x'_2,y'_2]=\nit{bl}_3$.

\ignore{\vspace{-7mm}
\begin{eqnarray*}
&&\hspace{0.2cm} \nit{Author}'[\nit{aid}_1, x_1,y_1]=\nit{bl}_2, \nit{Author}'[\nit{aid}_2, x_2,y_2]=\nit{bl}_2 \leftarrow \nit{Author}[\nit{aid}_1, x_1,y_1]=\nit{bl}_1 , \\&&
\hspace*{1.05cm}  \nit{Author}'[\nit{aid}_2,x_2,y_2]=\nit{bl}_2, \nit{PaperAuthor}'(\nit{pid}_1,\nit{aid}_1, x_1,y_1), \nit{NameSim}(x_1, x_2), \\&&\hspace*{1.05cm} \nit{PaperAuthor}'(\nit{pid}_2,\nit{aid}_2, x_2,y_2),
\nit{Paper}'[\nit{pid}_1,x_1,y_1,z_1,w_1,v_1]=\nit{bl}_3, \nit{bl}_1 < \nit{bl}_2,\\&&
\hspace*{7.05cm}  \nit{Paper}'[\nit{pid}_2,x_2,y_2,z_2,w_2,v_2]=\nit{bl}_3.
\end{eqnarray*}
\vspace{-7mm}
\begin{eqnarray*}
&& \hspace*{0.2cm}  \nit{Author}'[\nit{aid}_1, x_1,y_1]=\nit{bl}_2 \ , \nit{Author}'[\nit{aid}_2, x_2,y_2]=\nit{bl}_2 \leftarrow \nit{CoAuthor}'(\nit{aid}_1,\nit{aid}_3), \\ &&\hspace*{1.10cm}  \nit{CoAuthor}'(\nit{aid}_2,\nit{aid}_3), \nit{Author}'[\nit{aid}_1, x_1,y_1]=\nit{bl}_1 , \nit{Author}'[\nit{aid}_2, x_2,y_2]=\nit{bl}_2, \\ &&\hspace*{8.0cm}  \nit{NameSim}(x_1, x_2) \ , \nit{bl}_1 < \nit{bl}_2.
\end{eqnarray*}
\vspace{-2mm}   }

\item [{\bf 4.}] $\nit{Paper\mbox{{\it -}}OldVer}(\nit{pid},x,y,z,w,v,\nit{bl}_1) \  \leftarrow \ \nit{Paper}[\nit{pid},x,y,z,w,v] =\nit{bl}_1,$

\vspace{-1mm}
\hspace*{5.1cm}     $\nit{Paper}[\nit{pid},x,y,z,w,v]=\nit{bl}_2, \ \nit{bl}_1 < \nit{bl}_2.$

\item [{\bf 5.}] $\nit{Paper\mbox{{\it -}}Block}[\nit{pid},x,y,z,w,v]= \nit{bl} \;\;\leftarrow \;\;\nit{Paper}[\nit{pid},x,y,z,w,v]= \nit{bl},$

\vspace{-1mm}
\hfill $\nit{not} \ {\it Paper\mbox{{\it -}}OldVer}(\nit{pid},x,y,z,w,v, \nit{bl}).$
\end{itemize}
 By restricting the model of the program to attributes $\nit{PID}$ and $\nit{Block\#}$ of predicate $\nit{Paper\mbox{{\it -}}Block}$, we obtain blocks: $\{123, 205\}, \{195 ,769\}, \ldots$. That is, the papers with \nit{pids} $123$ and $205$ are blocked together;
similarly for those with \nit{pids} $195$ and $769$, etc.\boxtheorem
\end{example}
The execution of the blocking-program $\Pi^{\!\nit{Bl}}(D)$ will return in the end, for each entity-relation $R$ a list of subsets of the extension of $R$ in $D$. These subsets are blocks of $R$-records.
Pairs of records in a same block will be inputs to the classification model, which has to be independently constructed first.

\section{Classification Model Construction}\label{sec:MLClassification}

For both, the classification model construction and duplicate detection that uses it, weight-vectors for record-pairs have to be computed.
  The numerical values for these vectors come from features  related to similarity comparisons between attribute values for two records $r_1, r_2$.  Only a subset of record attributes are chosen, those attributes that have strong discriminatory power, to achieve maximum classification recall and precision (cf.~Section \ref{sec:features}).

\ignore{  The similarity functions $f_i$ explained, in Section \ref{sec:init}, are applied for obtaining  feature values for  each record-pair. In particular, weight-vectors \ $w(r_1,r_2) = \langle \cdots, w_i(\nit{f}_i(r_1[A_i],r_2[A_i])), \cdots \rangle$
are formed by applying predefined weights, $w_i$, on real-valued similarity functions, $f_i$, on pair of of values for attributes $A_i$. This is what similarity computation component does in {\em ERBlox}.}

The input to the SVM algorithm (that will produce the classification model) is a set of tuples of the form \ $\langle r_1, r_2, w(r_1,r_2), L\rangle$,  where $r_1,r_2$ are records (for the same entity) in the training dataset  $T$,  $L\in\{0,1\}$, and $w(r_1,r_2)$ is the computed weight-vector for the record-pair. In the {\em LogiQL} program, that input uses two defined predicates.
Predicate $\nit{TrainLabel}$ has two arguments: One
 for pairs of {\em rids}, $r_1r_2$, together, which is called ``the vector id" for vector $ w(r_1,r_2)=\langle w_1, \ldots, w_n\rangle$, and another to represents label $L$ associated to $w(r_1,r_2)$. Predicate $\nit{TrainVector}$ contains one argument for vector ids, and $n$ arguments to represent entries $w_i$ in the weight-vectors $w(r_1,r_2)$.

Several ML techniques are accessible from (or within) the {\em LogicBlox} platform, through the {\em BloxMLPack} library that provides a generic Datalog interface. Then, {\em ERBlox} can call  a SVM-based
classification model constructor,
through the general {\em LogiQL} program.

\ignore{
\comlb{OLD: Say how LogiQL interacts with SVM (or ML in general).}
\comlb{Is it really so? That SVM can be invoked from LogiQL? (this is my own question, not for the paper. Anyway, make what follows much more concrete and
showing the invocation of **SVM**, not in general.}
\comzb{Yes, SVM can be invoked from LogiQL.}  }

In particular, the {\em BloxMLPack}  wraps calls to the machine learning library in a
predicate-to-predicate mapping called {\em mlpack}, and manages marshalling the inputs and outputs to the machine
learning library from/to {\em LogiQL} predicates. This
is done via special  rules in {\em LogiQL} that come in two modes: the
learning mode (when a model is being learned, in our case, a SVM classification model), and the evaluation
mode (when the model is applied, for record-pair classification in our case) \cite{logicblox,Aref15}. We
do not give here the formal syntax and semantics for these rules, but just the gist by means of an example.

Assume that we want to train a SVM-model for \ent{Author}-record classification. For invoking SVM from {\em LogiQL}, a relation $\nit{InputMatrix}[j,i]$ is needed. It contains tabular data where each column ($j$) represents a
feature of \ent{Author}-records, while each row $i$ represents a vector id for which the tuple $\nit{TrainVector}(i, w_1, w_2,$ $ w_3)$ exists. So,  $\nit{InputMatrix}[j, i]$ represents the value of the feature $j$ in the weight-vector $i$. The following rules are used in {\em LogiQl} to populate relation $\nit{InputMatrix}$: \  (They involve predicates  $\nit{Feature}(``\nit{Fname}"), \ \nit{Feature}(``\nit{Lname}"),$ and $\nit{Feature}(``\nit{Affiliation}")$, associated to the three chosen attributes for  \ent{Author}-records. They appear in quotes, because they are constants, i.e.~attribute
names.)
\begin{eqnarray*}
&&\hspace*{-0.4cm}InputMatrix[``\nit{Fname}", i] = w_1, \ InputMatrix[``\nit{Lname}", i] = w_2, \\ &&\hspace*{0.8cm} \ InputMatrix[``\nit{Affiliation}", i] = w_3 \ \longleftarrow \ \nit{TrainVector}(i, w_1, w_2, w_3), \\ &&\hspace*{1.7cm} \nit{Feature}(``\nit{Fname}"), \ \nit{Feature}(``\nit{Lname}"),  \ \nit{Feature}(``\nit{Affiliation}").
\end{eqnarray*}
\noindent The following learning rule learns a SVM model for \ent{Author}, and stores the resulting model in the predicate $\nit{SVMsModel}(\nit{model})$:

\vspace{-0.50cm}

\begin{eqnarray*}
\nit{SVMsModel}(m) \ \longleftarrow \nit{mlpack} \ \ll  m=\nit{SVM}(\bar{p}), \nit{train} \gg \nit{InputMatrix}[j, i]=v, \\ && \hspace{-50mm} \nit{Feature}(j), \nit{TrainLabel}(i,l).
\end{eqnarray*}
Here, the head of the rule defines a predicate where the ML algorithm outputs its results, while
the body of the rule lists a collection of predicates that supplies data for the ML algorithm. In the above rule, the required parameters $\bar{p}$ for running the SVM algorithm are specified by the user.
The above rule is in the training mode.

\section{Duplicate Detection and MD-Based Merging}\label{Detection}

The input to the trained classifier is a set of tuples of the form $\langle r_1,r_2, w(r_1,r_2) \rangle$, where $r_1,r_2$ are record (ids) in a same
block for a relation $R$, and  $w(r_1,r_2)$ is the weight-vector for the record-pair
$\langle r_1, r_2\rangle$. The output  is a set of triples of the form $\langle r_1, r_2, 1\rangle$ or $\langle r_1, r_2, 0\rangle$.

Using  {\em LogiQL} rules, the triples $\langle r_1, r_2, 1\rangle$ form the extension of a defined predicate $R\mbox{-}\nit{Duplicate}$.

\begin{example}\label{ex.trainedClassifier}  (ex. \ref{LP} and \ref{ex:similarityComputation} cont.)
 Considering the previous \ent{Paper}-records, the input to the trained classifier consists of:  $\langle 123,$ $205, w(123,205) \rangle$, with $w(123,205) =[0.8, 1.0, 1.0, 0.7]$; and $\langle 195,769, w(195,$ $769)\rangle$, with  $w(195, 769)=[0.93, 1.0, 1.0, 0.5]$. 

In this case, the SVM-based classifier returns \ $\langle[0.8, 1.0, 1.0, 0.7],1\rangle$ and $\langle[0.93, 1.0, 1.0,$ $0.5],1\rangle$. Accordingly,  the tuples $\nit{Paper}\mbox{-}\nit{Duplicate}(123, 205)$ and $\nit{Paper}\mbox{-}\nit{Duplicate}(195,769)$ are created.
\boxtheorem
\end{example}
The extensions of predicates $R\mbox{-}\nit{Duplicate}$ will be the input to the merging process.

\ignore{\section{MD-Based Merging}\label{Merging}
\comlb{I moved here, at the beginning, what we had at the end of the the Overview section. From here ...}
\comlb{What follows is quite fuzzy. The obtention of a set of interaction-free MDs (or something that acts as such) has to be properly justified.}  }

Record merging is carried out through the enforcement of merge-MDs, as described in Section \ref{MLMDFramework}, where we showed that they form an {\em interaction-free} set. Consequently, there is a single
instance resulting from their enforcement. These MDs use application-dependent matching functions (MFs), and can be expressed by means of {\em LogiQL} rules.
Actually, the generic merge-MDs in (\ref{eq:offBmds}) can be expressed in their Datalog versions by means of  the above mentioned $R$\nit{-Duplicate} predicates.  The RHSs of MDs in (\ref{eq:offBmds}) have to be expressed
in terms of MFs, $\match_{A_i}$. All these become ingredients of a Datalog merge-program $\Pi^M$.

\begin{example} \label{ex:paperML}(ex. \ref{ex:similarityComputation} cont.) Duplicate \ent{Paper}-records are merged by enforcing the merge-MD:

$\nit{Paper}[\nit{pid}_1]\approx \nit{Paper}[\nit{pid}_2] \ \longrightarrow \ \nit{Paper}[\nit{Title},\nit{Year},
\nit{CID}, \nit{Keyword}] \doteq$\\
\hspace*{5.8cm} $\nit{Paper}[\nit{Title},$ $\nit{Year}, \nit{CID},$ $\nit{Keyword}]$.\boxtheorem
\end{example}

 The  general {\em LogiQL} program, $\Pi^M$, for  MD-based   merging contains rules as in {\bf 1.}-{\bf 4.} below:

\begin{itemize}
\item [{\bf 1.}]  The ground atoms of the form $R\mbox{-}\nit{Duplicate}(r_1,r_2)$ mentioned above, and those representing MFs, of the form $\match_{A}(a_1,a_2)=a_3$.

\item [{\bf 2.}]  For an MD \ $R[r_1] \approx R[r_2] \ \longrightarrow \ R[\bar{r}_1] \doteq R[\bar{r}_2]$,  the rules:
\ignore{\begin{eqnarray*}
&&R[r_1,\bar x_3]=\nit{bl}, \ R[r_2,\bar x_3]=\nit{bl} \ \longleftarrow \ \red{R\mbox{-}\nit{Duplicate}(r_1,r_2)}, \ R[r_1, \bar x_1]=\nit{bl},\\ &&\hspace*{6.35cm}  R[r_2, \bar x_2]=\nit{bl}, \ \match(\bar x_1,\bar x_2)=\bar x_3,
\end{eqnarray*}}
\begin{eqnarray*}
&&R(r_1,\bar x_3), \ R(r_2,\bar x_3) \ \longleftarrow \ R\mbox{-}\nit{Duplicate}(r_1,r_2), \ R(r_1, \bar x_1), \ R(r_2, \bar x_2),\\ &&\hspace*{4cm}  \ \match(\bar x_1,\bar x_2)=\bar x_3,
\end{eqnarray*}
where $\bar{x}_1, \bar{x}_2, \bar{x}_3$ stand  for all attributes of relation $R$, except for the {\em rid} and the block number (block numbers play no role in merging). $\match(\bar x_1,\bar x_2)=\bar x_3$ is just a shorthand to denote the componentwise application of $m$ individual MFs $\match_{A_i}$ \ (cf.~(\ref{eq:offBmds})).

At the end of the iterative application of these rules, there may be several tuples with different {\em rids} but identical ``tails". Only
one of those tuples is  kept in the resolved instance.


\item [{\bf 3.}]  As for the blocking-program $\Pi^{\!\nit{Bl}}(D)$ of Section \ref{MDBlocking}, we need rules specifying the old versions of a record:
\begin{eqnarray*} && R\mbox{-}\nit{OldVer}(r_1,\bar{x}_1) \ \longleftarrow \ R(r_1, \bar{x}_1), \ R(r_1, \bar{x}_2), \ \bar{x}_1 \prec \bar{x}_2,
\end{eqnarray*}
where $\bar{x}_1$ stands for all attributes other than {\em rid} and the block number; and $\bar{x}_1 \prec \bar{x}_2$ means componentwise comparison
of values according to the partial orders defined by the MFs. (Recall from Section \ref{sec:mds}, that each application of a MF makes us grow in the information lattice:
the highest values are the newest values.)

\item[{\bf 4.}] Finally, we introduce rules to collect, in a new predicate $R\mbox{-}\nit{ER}$,  the latest version of each record, to build the final resolved instance:
$$R\mbox{-}\nit{ER}(r,\bar{x}) \;\;\leftarrow \;\;R(r,\bar{x}), \ \nit{not} \ R\mbox{-}\nit{OldVer}(r,\bar{x}).$$
\end{itemize}

This is a stratified Datalog program that computes a single resolved instance in polynomial time in the size of the extensional database, in this case formed by
the contents of relations $R\mbox{-}\nit{Duplicate}$ and $D$.\footnote{As with the blocking-programs, the merge-programs can be obtained particularizing
the general programs in \cite{Bahmani12} to the case of interaction-free MDs \cite{BahmaniExten12}.}

\ignore{\comlb{I ignored below all the rubbish that was supposed to justify that the class of MDs is interaction-free: they are interaction free by syntactic construction!
Check if I missed anything relevant.} }

\ignore{
\red{Notice that the derived tables $R\mbox{-}\nit{Duplicate}$ that appear in the LHSs of the MDs (or in the bodies of the corresponding rules) are all computed before (and kept fixed during) the enforcement of the merge-MDs. Recall that we have atoms $R\mbox{-}\nit{Duplicate}(r_1,r_2)$ where $r_1,r_2$ are ids for records.  In particular, a duplicate relationship between any two records is not lost.
This has the effect of making the set of merging-MDs interaction-free, which results in a unique resolved instance.}
\red{As MD-based blocking, enforcing a set of blocking-MDs results in a unique instance, stratified
Datalog programs are expressive enough to express and enforce merging-MDs \cite{BahmaniExten12}.}
\red{A general matching
function that could potentially work for every attribute domain is a function that
treats attribute values as sets and takes the union of two sets whenever they need
to be identified \cite{Bertossi12}. To keep the ER instance accurate, and merge records without incurring loss of information, we use this matching function in Example \ref{ex:paperML}. When two \ent{Paper} records need to be identified, the union of the records is used, defined as the component-wise union of attribute values. Notice that we assume the attribute values are sets of
values.}
}

In our application to bibliographic datasets, we used as  matching functions ``the union case" \cite{BenjellounGMSWW09}, which was investigated in detail in \cite{Bertossi12} in terms of MDs. The idea is
to treat \ignore{non-numerical} attribute values as objects, i.e.~sets of pairs
attribute/value. For example, the address ``250 Hamilton Str., Peterbrook, K2J5G3" could be represented as the set $\{\langle {\sf number}, 250\rangle,$ $\langle {\sf stName}, \nit{Hamilton~Str.} \rangle,$ $\langle {\sf city}, \nit{Peterbrook}\rangle,$ $ \langle {\sf areaCode}, \nit{K2J5G3}\rangle\}$.
When two values of this kind are merged,
their union is computed. For example, the two strings ``250 Hamilton~Str., K2J5G3" and ``Hamilton~Str., Peterbook", represented as objects, are merged into ``250 Hamilton~Str., Peterbook,  K2J5G3"
\ \cite{Bertossi12}.
\red{ This generic merge function has the advantage that, in essence, the older pieces of information are preserved, and combined into a more complete value. In this example,
 the string ``250 Hamilton~Str., Peterbook,  K2J5G3" is more informative than the two strings initial strings, ``250 Hamilton~Str., K2J5G3" and ``Hamilton~Str., Peterbook".} In the
case of two alternative values, the two versions will be kept in the union, which may require some sort of domain-dependent postprocessing,
essentially making choices and possibly edits. In any case, working with the union case for matching dependencies is good enough for our purposes,
namely to compare traditional techniques with ours.

\ignore{\comRW{About the sentence ``This generic merge function has the advantage that information is preserved. It also works fine when two values complete each other.''\\
First, the expression ``works fine'' does not seem adequate for a formal discussion, authors can rephrase. Second, what do the authors mean by ``two values complete each other''?} }

\ignore{
\red{A general matching
function that could potentially work for every attribute domain is a
function that
treats attribute values as sets and takes the union of two sets whenever
they need
to be identified \cite{Bertossi12}.}
\red{ To keep the ER instance accurate, and
merge records without incurring loss of information, we use this matching
function in Example \ref{ex:paperML}. When two \ent{Paper} records need to
be identified, the union of the records is used, defined as the
component-wise union of attribute values. Notice that we assume the
attribute values are sets of
values.}}

We point out that MD-based merging takes care of ``transitive cases" produced by the classifier. More precisely,  if it returns  $\langle r_1, r_2,1\rangle$ and $\langle r_2, r_3,1\rangle$, but not $\langle r_1, r_3,1\rangle$, we still merge $r_1, r_3$ (even when $r_1 \approx r_3$  does not hold). Indeed, if  MD-enforcement first merges $r_1, r_2$ into the same record, the similarity between  $r_2$ and $r_3$ still holds (it was pre-computed and stored,
and not destroyed by the updating of attribute values of $r_2$). Then, the merge-MD will be applied to $r_3$ and the new version of $r_2$. Iteratively, $r_1, r_2, r_3$ will end up having the same attribute values (except for the {\em rid}).\footnote{Notice that there is certain similarity with the argument around the SFAI case of MDs in Section \ref{MDBlocking}. This not a coincidence: non-interacting MDs form a case of
SFAI, for any initial instance.}

There might be applications where we do {\em not} want this form of full entity resolution triggered by transitivity. If that is the case, we could use semantic constraints on the
ER result (or process). Actually, {\em negative rules} have been proposed in  \cite{BenjellounVLDB09}, and discussed in \cite{Bahmani12} in the context of general answer set programs
for MD-based ER. However, the introduction of constraints into Datalog changes the entire picture. Under a common approach, if the intended model of the program does not satisfy the constraint, it is rejected. This is not particularly appealing from the application point of view. An alternative is to transform constraints into non-stratified program rules, which would take us in general to the realm of ASPs \cite{brewka}. In any case, developing this case in full is outside the scope of this work.

\section{Experimental Results}\label{evaluation}

\begin{figure}[h]
  \hspace*{2.5cm}  \includegraphics[width=9cm]{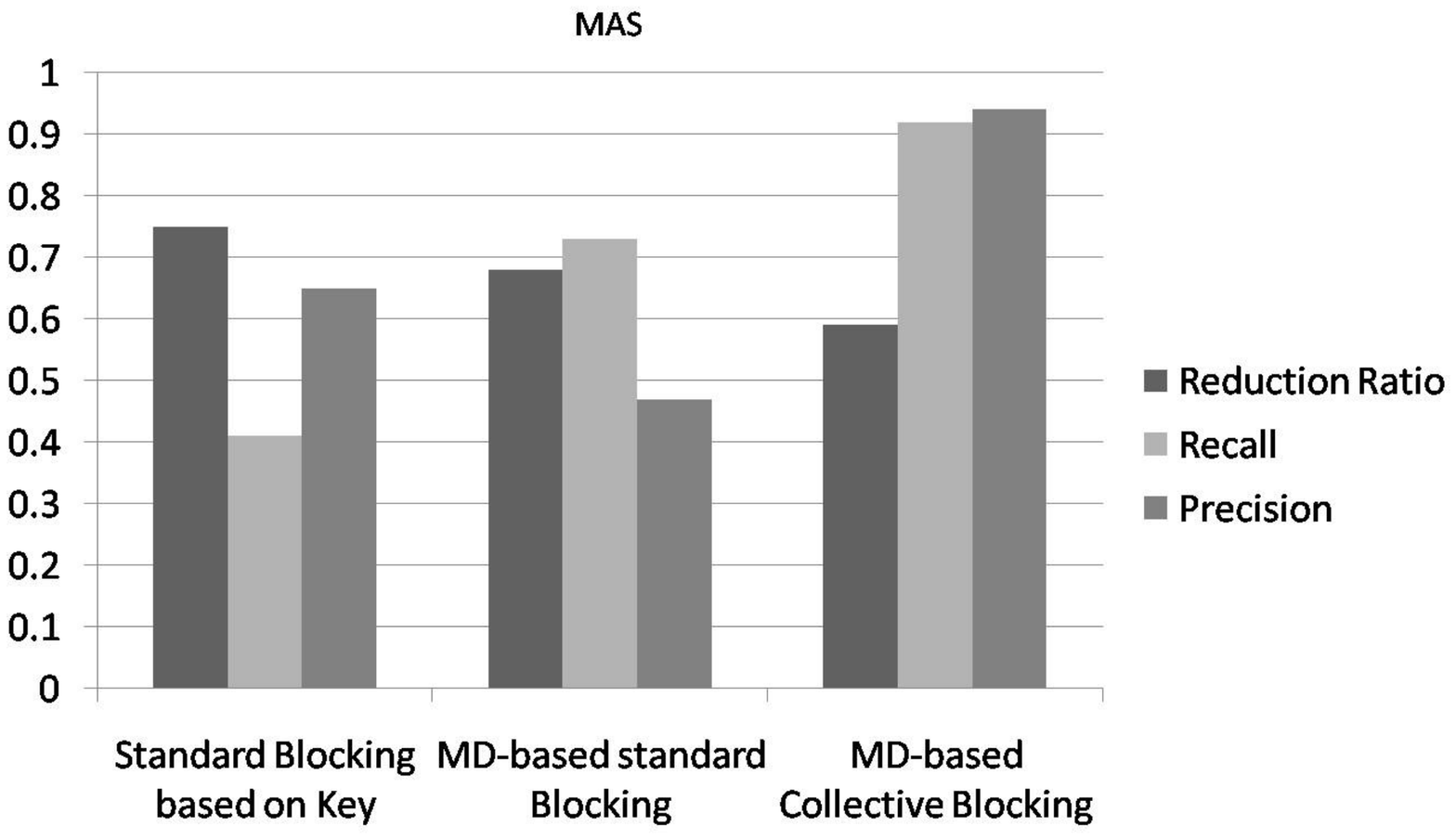}
  \vspace{-110pt}
  \caption{The experiments (MAS)}\label{fig:plot12}
 \end{figure}
\begin{figure}[h]
 \hspace*{2.5cm}   \includegraphics[width=8cm]{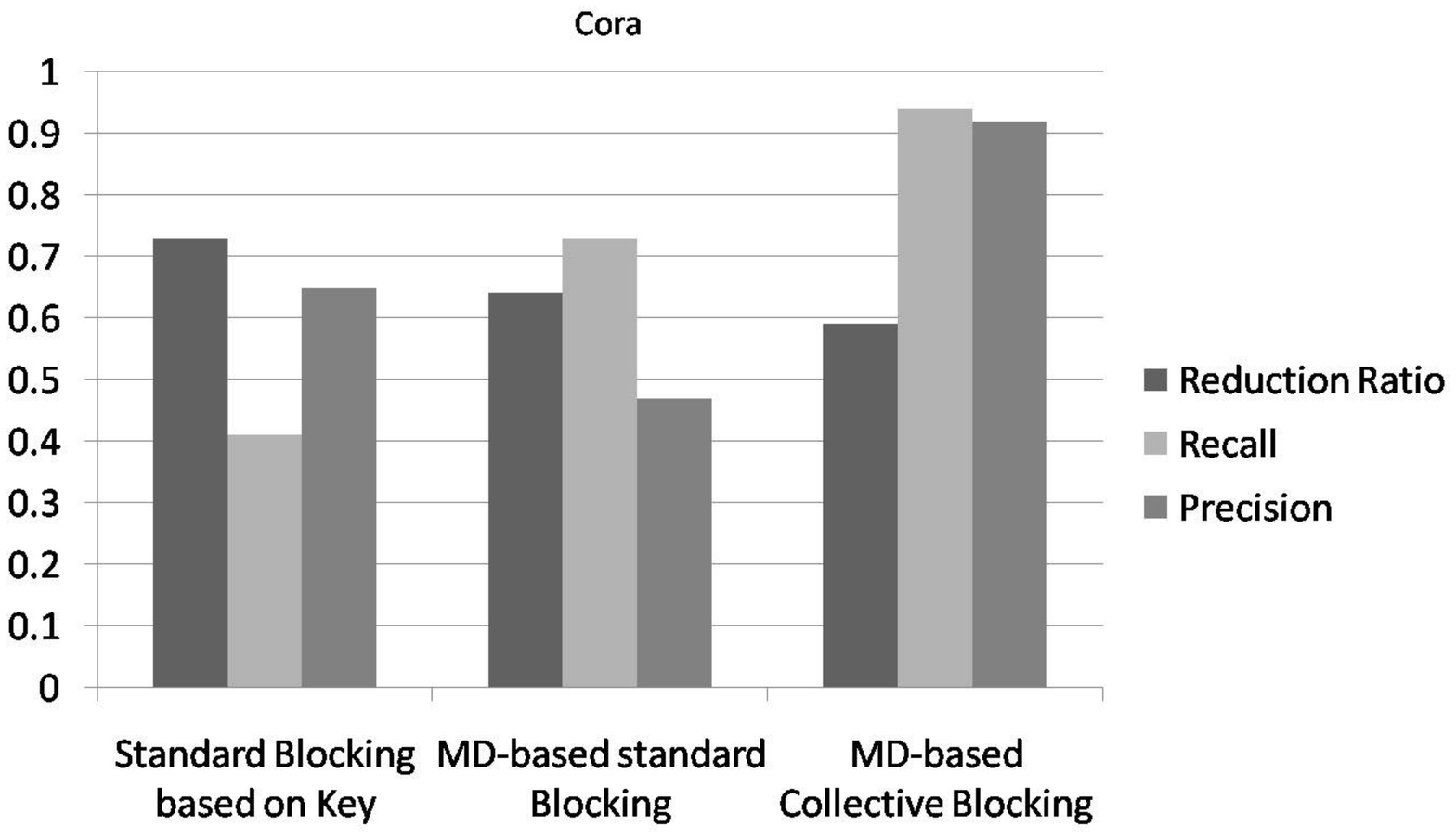}
  \vspace{-100pt}
  \caption{The experiments (Cora)}\label{fig:plot13}
   \end{figure}
\begin{figure}[h]
   \hspace*{2.5cm} \includegraphics[width=8cm]{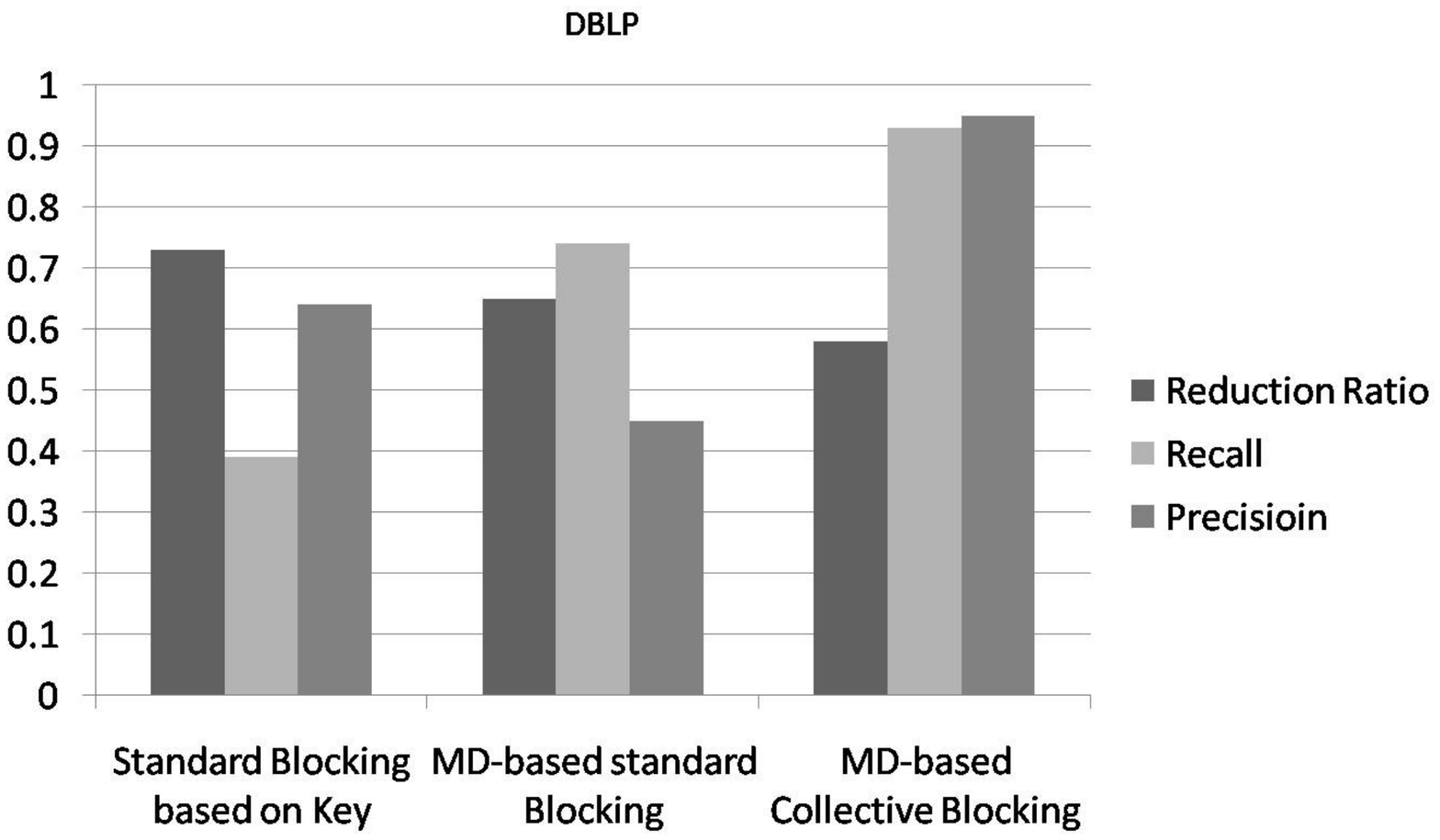}
  \vspace{-100pt}
  \caption{The experiments (DBLP)}\label{fig:plot11}
  \end{figure}

In comparison with {\em standard blocking} (SB) techniques, our experiments with the MAS dataset show that our approach to ER, in particular, through
the use of semantically rich  matching dependencies for blocking
result in lower {\em reduction ratio} for blocking, and higher {\em recall} and {\em precision} for classification.  These are positive results that can also be observed
in the experimental results with the DBLP and Cora Citation datasets. \ Cf.~ Figures \ref{fig:plot12}, \ref{fig:plot11}, and \ref{fig:plot13} (more details follow below).


We considered three different blocking techniques, shown, respectively, in the sets of columns in Figure~\ref{fig:plot12}: (a) {\em Standard Blocking} (SB), \ (b) MD-based Standard Blocking (MDSB), and (c) MD-based Collective Blocking
(MDCB), which we now describe:
\begin{itemize}
\item[(a)]
According to SB,
records are clustered
into a same block when they share the identical values for blocking keys \cite{jaro89}.

\item[(b)] MDSB generalizes standard blocking through the use of blocking-MDs  that consider on the LHS exactly the same attributes (actually, keys) as in SB. However,
 for some of the attributes, equality is
replaced by similarity, adding more flexibility and the possibility of increasing the number of two-record comparisons.

For example, the following could be an MD directly representing a blocking-key rule:
\begin{eqnarray*}
\nit{Author}(\nit{aid}_1, x_1,y_1, \nit{bl}_1) \ \wedge \ \nit{Author}(\nit{aid}_2, x_2,y_2, \nit{bl}_2) \ \wedge \ x_1 = x_2 \ \wedge\\ 
~~~~~~~~~~~~~~~ y_1 = y_2 \ \longrightarrow \  \ \nit{bl}_1 \doteq \nit{bl}_2;\hspace{-5mm}
\end{eqnarray*}
and the following could be a relaxed version of it, a single-relation MD where instead of equalities we now have similarities:
\begin{eqnarray*}
\nit{Author}(\nit{aid}_1, x_1,y_1, \nit{bl}_1) \ \wedge \ \nit{Author}(\nit{aid}_2, x_2,y_2, \nit{bl}_2) \ \wedge \ x_1 \approx_{\!\nit{Name}} x_2 \ \wedge \\ 
\hspace*{4cm} \ y_1\approx_{\!\nit{Aff}}y_2 \ \longrightarrow \  \ \nit{bl}_1 \doteq \nit{bl}_2.
\end{eqnarray*}
In this case we had as many MDs as blocking keys in SB, and they are each, single entity, such as (\ref{eqq:md1}) and (\ref{eqq:md2}) in Example
\ref{ex:blockLoqic}).

\item[(c)] MDCB uses, in addition to single-entity blocking-MDs, also
multi-relational MDs, such as (\ref{eqq:md3}) and
(\ref{eqq:md4}) in Example \ref{ex:blockLoqic}). In this case, the set of MDs contains all those in MDSB plus properly multi-relational
ones.
\end{itemize}

Reduction ratio refers to the record-blocking task of ER, and
is defined by $1- \frac{S}{N}$, where $S$ is
the number of candidate duplicate record-pairs produced by the
blocking technique, and $N$ is the total number of possible candidate
duplicate record-pairs in the entire dataset. If there are $n$ records
for an entity, then $N = n \times n$ for that entity.

 Reduction ratio
is the relative reduction in the number of candidate duplicate
record-pairs to be compared.
The higher the reduction ratio, the fewer the candidate
record-pairs that are generated, but the quality of the
generated candidate record-pairs is not considered
\cite{Christen11}. \ignore{Moreover, reduction ratio does not measure
the time taken for a particular implementation of a blocking
technique.}

That the
reduction ratio decreases from left  to right in Figure~\ref{fig:plot12} shows that the use of blocking-MDs increasingly captures more potential
record-pairs comparisons that would be missed otherwise.

\ignore{
 \begin{figure}[h]
 \begin{center}
 \includegraphics[width=4cm]{recPrec}
  \end{center}
 \vspace{-25pt}
  \caption{Precision and recall}\label{fig:PrecRec}
 \end{figure}
}

\bl{{\em Recall} and {\em precision} are measures of goodness of the result of the classification task \cite{Christen11}, in this case, of record-pairs as {\em duplicate} (i.e.~containing  duplicates of each other) or not.
  \ignore{They are both defined in terms of ratios of the sizes of the regions in Figure~\ref{fig:PrecRec}.}  Recall is defined by $\mbox{\it true positives}/(\mbox{\it true
positives} \ + \ \mbox{\it false negatives})$, whereas precision is defined by \
$\mbox{\it true positives}/$ $(\mbox{\it true
positives} \ + \ \mbox{\it false positives})$.}

\bl{Higher recall means  more true (i.e.~duplicate) candidate
record-pairs have been actually found; and  higher precision, that
more of the retrieved candidate duplicate record-pairs are
actually true.}
Since we want high recall and precision, then we prefer
a blocking technique that generates a small number of candidates for false
positives and false negatives.

\vspace{-3cm}

Our experiments focused mainly on the recall and precision of the
overall results after classification (and before
merging). They indirectly allows for the evaluation of the blocking techniques, as well.
Actually, recall measures the effectiveness a blocking technique through non-dismissal
of true candidate duplicate record-pairs. Similarly,
 a high precision reflects that the blocking
technique generates mostly true candidate
duplicate record-pairs. Inversely, a low precision shows
a large number of non-duplicate record-pairs is also considered, through blocking, as
candidate duplicate record-pairs.
\ignore{This is because duplicate record-pairs detected by the
classifier were first candidate duplicate record-pairs obtained from
blocking. As a consequence, true duplicate record-pairs from
classifier correspond to true candidate duplicate record-pairs
from blocking. In other words, if the record blocks are too
small, resulting in fewer number of candidate duplicate record-pairs,
then true duplicate record pairs may be missed, then reducing precision and recall of results after classification.}
We can see that it becomes crucial
to verify that filtering out record-pairs by a particular blocking technique
does not affect the quality of the results obtained after classification.

All the three above mentioned  measures were computed by cross-validation,
on the basis of the training data. Approximately 70\% of the training data was
used for training, and the other 30\%,  for testing. The MAS dataset includes 250K authors, 2.5M papers,
and a training set. For the authors dataset, the training and test
sets contain 3,739 and  2,244 cases (author ids), respectively.
Figures \ref{fig:plot12}, \ref{fig:plot11} and \ref{fig:plot13}
show the comparative performances of {\em ERBlox} with the three forms of blocking mentioned above, for three different
datasets.
In all cases, the same SVM technique was applied.

\vspace{-3cm}

In our concrete application domain,
standard blocking based on key-equalities of
\ent{Paper}-records of the MAS dataset used attributes \nit{Title},
\nit{Publication Year},  and \nit{Conference ID}, together, as one blocking
key. The MD-version of this key, for MD-based standard
blocking and MD-based collective blocking, is the MD
(\ref{eqq:md1}) in Example \ref{ex:blockLoqic}. According to it, if two
records have similar titles, with the same publication year and
conference ID, they have the same block numbers.
Deciding which attribute equalities become similarities is domain-dependent.

\ignore{\red{By a domain expert?s knowledge, we could decide which of the
attributes, appeared as keys in the standard blocking based on
key-equality, should be expressed by similarity or equality atoms in
$\nit{LHSs}$ of blocking-MDs.} For example,
for standard blocking based on key-equalities of
\ent{Paper}-records of the MAS dataset, we use attributes title,
publication year and conference ID together, as one blocking
key. The associated MD for this key, in the MD-based standard
blocking and MD-based collective blocking, is the MD
(\ref{eqq:md1}) in Example \ref{ex:blockLoqic}, where if two
records have similar titles with the same publication year and
conference ID, they have same block numbers. This means that we need to
check the similarity of papers titles, and equalities of  their publication
years and conference IDs to put the papers in the same block.
}

Standard blocking based on key-equalities has higher
reduction ratio than MD-based standard blocking, i.e.~the former generates fewer candidate duplicate record-pairs.
Standard blocking also leads to higher precision than MD-based
standard blocking, i.e.~we can trust more
candidate duplicate record-pairs judgements obtained via
standard blocking. However, this standard blocking
is very conservative, and has a very low rate of
recall, i.e.~many of the true candidate duplicate
record-pairs are not identified as such. All this makes sense
since with standard blocking based we only
consider equalities of blocking keys, not similarities.

Precision and recall of MD-based collective blocking are
higher than the two standard blocking techniques. This
emphasizes the importance of MDs that support collective
blocking, and shows that blocking based on string similarity
alone fails to capture the semantic interrelationships that naturally hold in
the data. On the other side, MD-based collective blocking has lower
reduction ratio than standard MD-based blocking, which may lead to better ER results, but may impact
computational cost: larger blocks may be produced, and then, more candidate
duplicate record-pairs become inputs for the classifier.
In blocking, this is a common tradeoff that needs to be
considered \cite{Christen11}. \ignore{On the one hand, having a large number of smaller
blocks will result in fewer candidate duplicate record pairs
that will be generated, probably increasing the number of true
duplicate record pairs that are missed. On the other hand,
blocking techniques that result in larger blocks generate a
higher number of candidate duplicate record pairs that will
likely cover more true duplicate pairs, at the cost of having
to compare more candidate pairs }

\ignore{
\vspace{8mm} Notice that that different sets of blocking-MDs
for MD-based collective blocking have different impact on
reduction ratio, precision and recall, so as standard blocking
based on key-equality depends on the choice of blocking keys. }

Overall, the quality of MD-based collective blocking dominates
standard blocking, both in its key-based and MD-based forms,  for the three datasets.

\section{Related Work}\label{RelWork}

An unsupervised clustering-based approach to collective deduplication is proposed in \cite{Bhattacharya}. While traditional deduplication techniques assume that only similarities between attribute values are available, in relational data the entities are assumed to have additional relational information that can be used to improve the deduplication process. This approach falls in the context of {\em relational learning} \cite{getoor}.
More precisely, in \cite{Bhattacharya}, a relationship graph is built whose nodes are the entities (records), and edges indicate entities which
co-occur. The graph supports  the propagation of similarity information  to related entities. In particular, the similarity between two nodes is calculated as the weighted sum of the attribute-value similarity and their relational similarity (as captured through the graph). Experimental results \cite{Bhattacharya} show that this form of {\em collective  deduplication} outperforms traditional deduplication.

 \red{The  approach to ER in \cite{Bhattacharya} could be seen as implicitly involving collective blocking, where relationships
between entities and similarities between attribute values are used to create the blocks of records. However,  this form of collective blocking does not take advantage of a declarative, logic-based semantics. In contrast, a {\em relationship graph} is used for collective deduplication. In our case,  semantic information for this task is captured by matching dependencies. Most importantly, the main focus of our approach to ER is MD-based collective blocking. For this reason, our experiments compare this approach with other blocking techniques. A comparison of our whole approach to ER with other (whole) collective approaches to ER, such as that in \cite{Bhattacharya} has to be left for future research. However, the results of such a comparison may not be very eloquent, because our approach is based on crucial intermediate techniques, such as the use of SVM for the classification task, which is
somehow orthogonal to the blocking approach. }

\ignore{
\comRW{Authors discuss cluster-based collective deduplication and say that it outperforms traditional deduplication. I wonder: How does this technique compare against standard blocking? Why didn't the authors compare the proposal against this method that seems to be one of the best for the job?}
}

 {\em Dedupalog}, a declarative approach to collective entity deduplication in the presence of constraints, is proposed in \cite{Arasu09}. Constraints are represented by a form of Datalog language. The focus of this work is
 unsupervised clustering, where constraints are an additional element. Clusters of records make their elements candidates for merging, but blocking
 {\em per se} or the actual merging are not main objectives.  However, this kind of clustering could be interpreted as a form of blocking. The additional use of constraints could be seen as a form
 of {\em collective clustering}. In \cite{Arasu09}, equality-generating dependencies were used as hard constraints, and clustering-rules as weak constraints.


 Our approach can also be seen as a form of relational learning. However, in our case, the semantic relational information (constraints) are, in some sense, implicitly captured through matching dependencies. Their semantics is non-classical (it is chase-based
 as seen in Section \ref{sec:mds}), and involves directly the blocking or merging processes, as opposed to having higher-level logical constraints ``compiled" into them.
In our case, the proper learning part of the process, i.e.~classification-model learning via SVM, is supervised,\footnote{We refer to \cite{Kopcke08} for a discussion on supervised vs. unsupervised approaches.} but it does not use any kind of additional relational knowledge. In this regard, it is worth pointing out to quite recent research proposing  supervised ML-techniques for classification that involve
semantic knowledge in the form of logical formulas in kernels for kernel-based methods (such as SVM) \cite{diligenti}.

Various blocking techniques have been proposed,  investigated and  applied. See  \cite{Baxter03,Christen11,Draisbach, Papadakis, surveyBlocking} for  comprehensive surveys and comparative studies.
To the best of our knowledge, existing approaches to blocking
are inflexible and limited in that they:
(a) allow blocking on only single entity types, in isolation from other entity types, or (b) do not take advantage of valuable domain or semantic knowledge.
\bl{Possible exceptions are \cite{Nin07,Rastgoi11}.
Collective blocking in \cite{Nin07}  disregards blocking keys and creates
blocks by considering exclusively the relationships between entities. The relationships correspond to links in a graph connecting entities, and blocks are formed by groping together
entities within neighborhoods with a predefined (path) ``diameter". Under this approach, in contrast with ours (cf.~Example \ref{ex:IntroBlock}), relationships are not declarative, and blocking decisions on one  entity do not have a direct, explicit  impact on blocking decisions to be made on another related entity.}


In \cite{Rastgoi11},  similarity of blocking keys and relational relationships are considered for blocking in the context of identification of duplicates (not the merging). However,
the semantics of relational relationships (or closeness) between blocking keys and entities is not fully developed.


\ignore{To our knowledge, the only work that briefly applies semantic knowledge for blocking is .  The framework is proposed for a large-scale collective entity matching that applies similarity of blocking keys and relational closeness for blocking of entities.  However, the authors do not offer precise semantics for capturing relational closeness between entities.  Relational closeness is incorporated in an ad-hoc way. As a collective deduplication framework, it lacks a merging component to make a single representation from the duplicate entities.}


{

\section{Conclusions}\label{conclude}

We have shown that matching dependencies, a new class of semantic constraints for data quality and cleaning, can be profitably integrated with traditional
ML-methods, in our case for developing classification models for entity resolution. These dependencies play a role not only in their intended goal, that of merging duplicate representations, but also in the
record-blocking process that precedes the proper learning task.  At that stage they declaratively capture semantic information that can be used to enrich the blocking activity.

MDs declaration and enforcement, data processing in general, and machine learning  can all  be integrated using the {\em LogiQL} language.
Actually, all the data extraction, movement and transformation tasks
are carried out via {\em LogiQL}, a form of extended Datalog supported by the {\em LogicBlox} platform.

In this regards it is interesting to mention that  Datalog has been around since the early 80s,
as a declarative and executable rule-based language for relational databases.
It was used mostly in DB research, until recently. In the last few years Datalog
 has experienced a revival, and many new applications have been found.

{\em LogiQL}, in particular, is being extended in such a way it can smoothly interact with optimization
and machine learning algorithms, on top of  a single platform.
Data for optimization and ML problems stored as ``extensions" for a relational database (that is a component of
{\em LogicBlox}), and
Datalog predicates. The results of those algorithms can be automatically stored in existing  database predicates or  newly defined Datalog predicates,
for additional computations or query answering.  Currently new ML methods are being implemented as components of  the {\em LogicBlox} system \ (cf.~Figure~\ref{fig:logiql}).

  \begin{figure}
  \begin{center}
    \includegraphics[width=12.5cm]{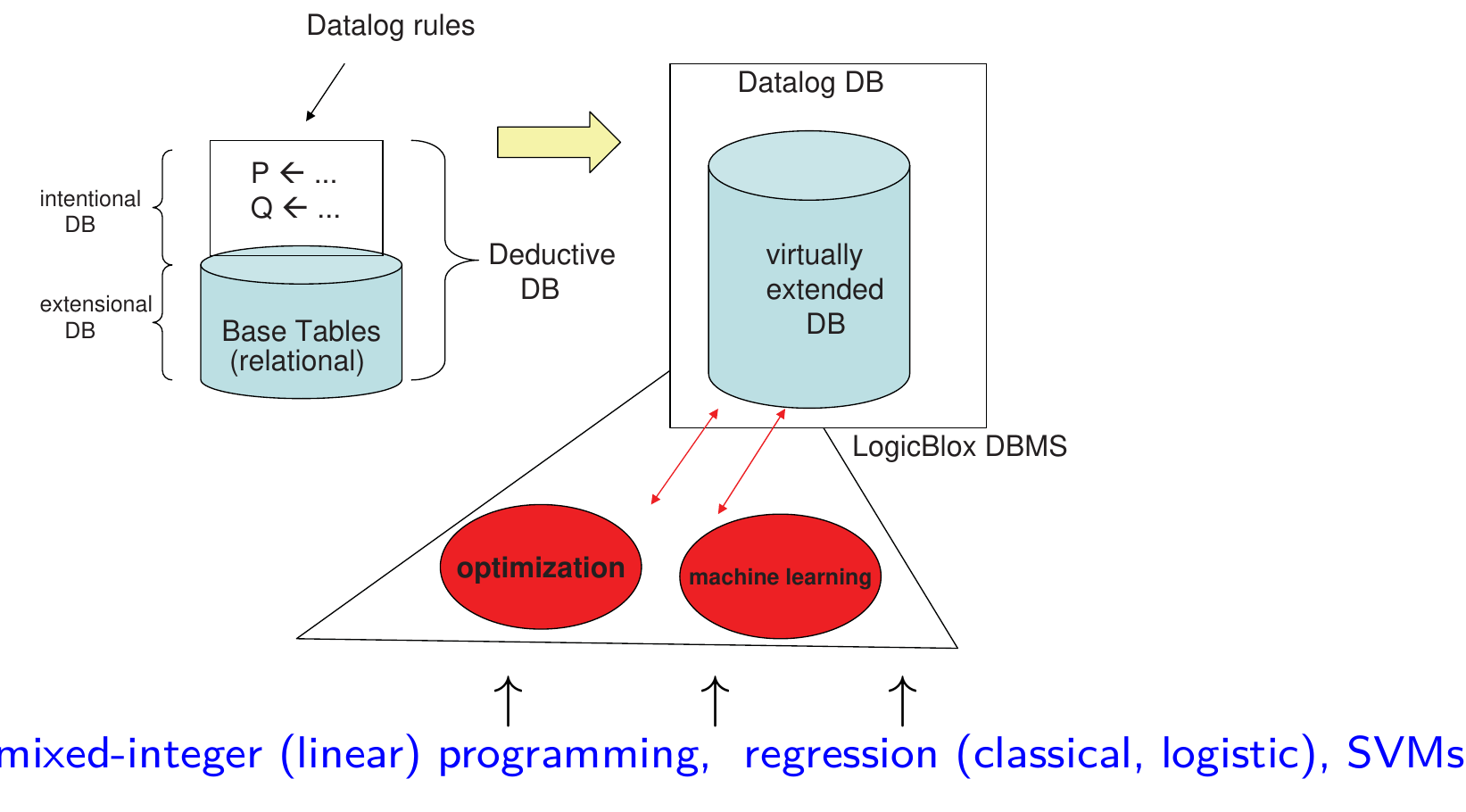}
  \end{center}
  \vspace{-9mm}
  \caption{{\em LogiQL} and extended {\em LogicBlox}}\label{fig:logiql}
   \end{figure}

Our work can be extended in several directions, some of which have been mentioned in previous sections. A most interesting extension would consider the use of more expressive blocking-MDs than those of the
form \ref{eq:newMDs}. Actually, they could have in their RHSs attributes other than \nit{Bl\#}, the block attributes. As a consequence, blocking-MDs, together with making block numbers identical, would make
identical pairs of application-dependent values for
some other attributes.
 Doing this would refine the blocking process itself (modifying the data for the next applications of blocking-MDs), but would also prepare \ignore{This affects the {\em ERblox} in two ways. First, updating other attributes values, as result of blocking-MDs enforcements, helps to block the records with higher precision and recall. Second, if we change the attribute values during the blocking, then we enrich} the data for the next task, that of classification for entity resolution. \ignore{Consequently, we would have higher precision and recall for duplicate detection since we deal with more complete information.  None of the existing blocking techniques aims at this point. However, in this case, enforcing the blocking-MDs results in more than one unique instance $D^{\nit{Bl}}$.}

\ignore{\red{Furthermore, it is possible to perform MD-based blocking iteratively as proposed in \cite{Garcia-Molina09}. This means that the results of the  MD-based merging are reflected to subsequently processed blocks. Blocks are now iteratively processed until no block contains any more matching records.}}

\vspace{1mm}\noindent {\small {\bf Acknowledgments:} \ Part of this research was funded by NSERC Discovery Grant \#250279-2011, and
the NSERC Strategic Network on Business Intelligence (BIN). \ Z. Bahmani and L. Bertossi are very much grateful for the
support from LogicBlox during their internship and sabbatical visit. }


\appendix

\section{\red{Relational MDs and the UCI Property}}\label{sec:relMDs}

Here, we formally extend the class of  matching dependencies (MDs) introduced in Section \ref{sec:mds}, which we will call  {\em classical MDs}, to the larger class of  {\em relational MDs}. This extension is motivated by
the application of MDs to blocking for entity resolution, but applications can be easily foreseen in other areas where declarative relational knowledge may be useful in combination with matching and merging.

We also identify classes of relational MDs for which a single clean instance exists, no matter how the MDs are enforced, that can be computed through the chase procedure in polynomial time in the size of the database on which the MDs are enforced. We say that the MDs (in some cases in combination with an initial instance) have the {\em unique clean instance property} (UCI property). More details can be  found in  \cite{Bertossi12,BahmaniExten12,corrFlairs17}.

\ignore{
In MDs, tuples for different relations may be related via
attributes in common. The way attribute values in tuples in certain relations are merged, as a result of enforcing an MD, may influence the way attribute values for tuples in  other relations are merged. For capturing this, we extend the class of classical MDs to the class of {\em relational MDs}, where semantic information is used to express relationships between different relations and their corresponding similarity conditions.

More formally, {\em relational MDs} are formulas of the form:
}

\begin{definition} \label{def:relMD} \em Given a relational schema $\mathcal{R}$, a  {\em relational MD} is a  formula of the form:
\begin{eqnarray}
\varphi\!:  \ \ \forall t_1 t_2 \bar{t}_3 \ \forall y_1 y_2  \bar{x}_1 \bar{x}_2 \bar{x} \ (R_1(t_1, y_1, \bar{x}_1) \ \wedge \   R_2(t_2, y_2, \bar{x}_2)  \ \wedge \  \psi(\bar{t}_3, \bar{z}) \nonumber \\  \hspace*{4cm} \ \longrightarrow  \ \ y_1 \doteq y_2). \label{eq:123}
\end{eqnarray}
Here, $R_1, R_2 \in \mathcal{R}$, the $\bar{x}_i$, etc. are lists of variables, and the $y_i$ are single variables, the $t_i$ are  tid variables, and the $\bar{t}_i$ are lists of tid variables. $R_1(t_1,y_1, \bar{x}_1)$,  $R_2(t_2, y_2, \bar{x}_2)$, are the {\em leading atoms}. Formula $\psi(\bar{t}_3, \bar{z})$ is a conjunction of similarity atoms and relational atoms (with predicates in $\mc{R}$), with $\bar{z}\cap (\bar{x}_1 \cup \{y_1\})\neq \emptyset$, $\bar{z}\cap (\bar{x}_2 \cup \{y_2 \})\neq \emptyset$.  \boxtheorem
\end{definition}
It is worth comparing classical MDs in (\ref{eq:md12145}) with this extended form. Here, the arguments in the {\em relational part} of the MD, namely in   $\psi(\bar{t}_3, \bar{z})$, interact via variables in common (joins) with the arguments in the leading  atoms.

\begin{example}\label{ex:relationalMD1}  (ex. \ref{ex:IntroBlock} cont.) \label{ex:block+}\ For schema $\nit{Author(Name,Aff,PapTitle,Bl\#)},$ $ \nit{Paper(PTitle,Kwd, Venue,Bl\#)}$, the following is a (properly) relational MD:

\begin{eqnarray}
\varphi\!: \ \underline{\nit{Author}(t_1,x_1,y_1, p_1,\nit{bl}_1)} \ \wedge \  \underline{\nit{Author}(t_2,x_2,y_2, p_2,\nit{bl}_2)}  \ \wedge \  x_1 \approx x_2  \ \wedge \nonumber \\   \nit{Paper}(t_3,p_1',z_1, w_1, \nit{bl}_4) \ \wedge \
 \nit{Paper}(t_4,p_2',z_2, w_2, \nit{bl}_4)  \ \wedge \ p_1 \approx p_1' \ \wedge \ p_2 \approx p_2' \nonumber \\ \longrightarrow \  \nit{bl}_1 \doteq \nit{bl}_2.\hspace{1cm} \label{eq:blmd2}
 \end{eqnarray}
Here, the leading atoms are underlined. They contain the two variables that appear in the identification atom on the RHS. Notice that there is an implicit similarity atom (an equality) represented by
 the use of the shared (join) variable $\nit{bl}_4$.\boxtheorem
\end{example}
The chase-based semantics developed for classical MDs can be applied to relational MDs, without any relevant change: the new relational conditions on the RHSs have to be made true when MDs are enforced.

On the classical side of MDs (cf.~Section \ref{sec:mds}), two special classes of  MDs were identified: similarity-preserving MDs, and interaction-free (IF) MDs. They have the UCI property.
On the relational side of MDs, similarity-preserving MDs (i.e.~that use similarity-preserving matching functions) are clearly UCI, because only new additional conditions  have to be verified before enforcing
an MD.
 We  proceed now to  generalize  the interaction-free class to the relational case, and prepare the ground for introducing a new class of relational MDs, the SFAI class.

\begin{definition}\label{def:relationMDIF} \em (a) For a relational MD $\varphi$, $\nit{ALHS}(\varphi)$ denotes the sets of attributes (with their predicates) appearing in similarity atoms in its LHS. \ $\nit{ARHS}(\varphi)$
denotes the set of attributes appearing  in the identity atom (with $\doteq$) in  its RHS.\footnote{We are making a distinction with $\nit{LHS}(\varphi)$ and $\nit{RHS}(\varphi)$ that denote the set of {\em atoms} in the
LHS and RHS side of $\varphi$, respectively.} (Notice from (\ref{eq:123}) that variables $y_1, y_2$ in the RHS have implicit predicates, say $R_1[Y_1], R_2[Y_2]$.)


\noindent (b) A set of relational MDs $\Sigma$ is  {\em interaction-free} (IF) if, for every $\varphi_1, \varphi_2 \in \Sigma$, $\nit{ALHS}(\varphi_1) \cap \nit{ARHS}(\varphi_2)=\emptyset$. Here,
$\varphi_1$ and $\varphi_2$ can be the same.\boxtheorem
\end{definition}

In Example \ref{ex:relationalMD1}, $\nit{ALHS}(\varphi)= \{ \nit{Author}[\nit{Name}], \nit{Author}[\nit{PTitle}], \nit{Paper}[\nit{PTitle}],$ $ \nit{Paper}[\nit{Bl\#}] \}$ and $\nit{ARHS}(\varphi)= \{\nit{Author}[\nit{Bl\#}] \}$. Since $\nit{ALHS}(\varphi) \cap \nit{ARHS}(\varphi) = \emptyset$, \ $\Sigma = \{\varphi\}$ is IF. 

For the same reasons as for similarity-preserving relational MDs, enforcing IF sets of relational MDs on an initial  instance results in a single clean instance that can be computed in polynomial time in the size of the initial instance. Accordingly, set of IF relational MDs have the UCI property.

The  class of relational MDs we will introduce next requires its combination with the initial instance.
\ignore{The new class we now introduce does. It is that of {\em SFAI combinations} of relational MDs and initial instances. Notice that this class has been identified as relevant due to the applications of MDs in this work.}

\begin{definition}\label{def:SFAInew}
\em Let $\Sigma$ be a set of relational MDs and $D$ an initial instance. The combination of $\Sigma$ and $D$ is {\em similarity-free attribute intersection} (sometimes we simply say that $(\Sigma, D)$  is SFAI)  if one of the following holds:
(below $\varphi_1, \varphi_2$ can be the same)

 \noindent
(a) \ There are  no $\varphi_1, \varphi_2\!\in\!\Sigma$, with  $\nit{ALHS}(\varphi_2) \cap \nit{ARHS}(\varphi_1)\neq \emptyset$, i.e., $\Sigma$ is {\em interaction-free}.

\noindent (b) \  For every $\varphi_1, \varphi_2 \in \Sigma$ and attribute $R[A] \in \nit{ALHS}(\varphi_2) \cap
\nit{ARHS}(\varphi_1)$,  it holds: If $S_1, S_2 \subseteq D$ with $R(\bar{c}) \in  S_1 \cap S_2$, then  $\nit{LHS}(\varphi_1)$ is not true in $S_1$ or $\nit{LHS}(\varphi_2)$ is not true in $S_2$.\footnote{We informally say that $\varphi_1$ is not applicable in $S_1$, etc.} \boxtheorem
\end{definition}

In condition (b) above,  $S_1$ and $S_2$ could be the same. Notice that condition (b) is checked only against the initial instance, and not on later instances obtained along a chase sequence. Since the SFAI notion depends on instances, we consider SFAI to be a semantic class,
as opposed to the two syntactic ones we considered before in this section.

In general, different orders of MD enforcements  may result in different clean instances, because tuple similarities may be broken  during the chase with interacting MDs and non-similarity-preserving MFs, without reappearing again \cite{Bertossi12}.
Intuitively, with SFAI combinations, two similar tuples in the original instance $D$ -or becoming similar along a chase sequence- may have the similarities broken
in a chase sequence, but they will   reappear later on in the same and the other chase sequences. Thus, different orders of MD enforcements cannot lead in the end to different clean instances.
This behavior can be better appreciated in Example \ref{ex:newclass1} below.

As expected, the notion of SFAI class  can be applied to classical MDs. Notice that for a classical MD, the LHS of an MD  is verified against pairs of tuples from the instance. Thus, for a set of classical MDs, $S_1$ and $S_2$ in condition (b) of Definition \ref{def:SFAInew} take the form $\{t_1, t_2\}$ and $\{t_2, t_3\}$, respectively.

\begin{remark}\label{remarSFAI}  A combination formed by a set of classical MDs $\Sigma$ and an instance $D$ is SFAI if there are  no $\varphi_1, \varphi_2\!\in\!\Sigma$, with $\nit{ALHS}(\varphi_2) \cap \nit{ARHS}(\varphi_1)\neq \emptyset$, or,  otherwise, for every $\varphi_1, \varphi_2 \in \Sigma$ and attribute $R[A] \in \nit{ALHS}(\varphi_2) \cap
\nit{ARHS}(\varphi_1)$ it holds: If  $t_1,t_2, t_3 \in D$, then $\nit{LHS}(\varphi_1)$ is not true in $\{t_1, t_2\}$ or $\nit{LHS}(\varphi_2)$ is not true in $\{t_2,t_3\}$. \boxtheorem
\end{remark}

\begin{example}  \label{ex:newclass1}   Consider predicate $R(A,B,C)$, the instance $D_0$, and the set of classical MDs $\Sigma$ below:


{\small
\begin{multicols}{2}

\begin{center}
$\varphi_1: R\left[A\right] \approx R\left[A\right] \rightarrow R\left[B\right] \doteq R\left[B\right]$,\\
$\varphi_2: R\left[B\right] \approx R\left[B\right] \rightarrow R\left[C\right] \doteq R\left[C\right]$.
\end{center}

\begin{center}
\begin{tabular}{c|c|c|c|}\hline
$R(D_0)$&$A$ & $B$ & $C$\\ \hline
$t_1$&$a_1$ & $b_1$ & $c_1$\\
$t_2$&$a_2$ & $b_2$ & $c_2$\\
$t_3$&$a_3$ & $b_3$ & $c_3$\\ \cline{2-4}
\end{tabular}
\end{center}
\end{multicols} }

Sometimes we use tids to denote a whole tuple (or record): if $t$ is a tuple identifier in instance $D$, $\bar{t}$ denotes the tuple in $D$ identified by $t$: $\bar{t} =R(c_1,\ldots, c_n)$.
If $\cal{A}$ is a sublist of the attributes of predicate $R$, then $t[\mc{A}]$ denotes the restriction of $\bar{t}$ to $\cal{A}$.

$\Sigma$ is interacting (i.e.~not IF), because  $\nit{ARHS}(\varphi_1) \cap \nit{ALHS}(\varphi_2)= \{R[B]\}$. 
Assume now that the only similarities that hold in the data domain $U$ are $a_1\approx a_2$,  $a_1\approx a_3$ and $b_3 \approx b_4$, with $b_4 \in \nit{Dom}(B) \smallsetminus \nit{Adom}(D_0)$.

Since $\varphi_2$ is not applicable in $D_0$ (i.e., its LHS is not true), $(\Sigma,D_0)$ is SFAI. Notice that $b_3 \approx b_4$ does not matter, because there is no tuple in $D_0$ with $b_4$ as value for $R[B]$. If we had $b_2\approx b_3$, with $t_2[B]=b_2, t_3[B]=b_3$ in $D_0$, $\nit{LHS}(\varphi_1)$ would be true in $\{t_1, t_2\}$, and $\nit{LHS}(\varphi_2)$ would be  true in $\{t_2,t_3\}$.

We will  show that the enforcement of $\Sigma$ on $D_0$  generates  a unique clean instance, through different chase sequences.  First, we show a possibly chase sequence
$D_0, D_1,D_2,D_3,D_4, D_5, D_6$, with $D_6$ a stable instance.  The
matching functions are as follows:

{\small \begin{center}
 $M_B(b_1, b_2)= b_{12}$,\ \ $M_B(b_2, b_3)= b_{23}$, \ \ $M_B(b_{12}, b_{123})= b_{123}$, \  \ $M_B(b_{12}, b_{3})= b_{123},$ \\ $M_C(c_1, c_2)=c_{12}$,\  \  $M_C(c_2, c_{3})=c_{23}$,\  \  $M_C(c_{12}, c_{3})=c_{123},$ \ \ \ \ $M_C(c_{12}, c_{123})=c_{123}$. \end{center}}

\noindent As a result of enforcing $\varphi_1$ on $D_0$ first, the  tuples $t_1$, $t_2$ get the identical values for $R[B]$, as shown in the new instance $D_1$ (cf.~Figure~\ref{fig:d1}). Next, since
$t_1$ and $t_2$ have same value for $R[B]$, we
can enforce $\varphi_2$, leading to $t_1, t_2$ getting the same value for $R[C]$, as shown
in instance $D_2$ (cf.~Figure~\ref{fig:d1}). As we can see, through  MD
enforcement new similarities may be created, in this case $t_1[B] =t_2[B]$ in $D_1$.  Furthermore, the equality of values for
attribute $R[B]$ feeds the LHS of $\varphi_2$.

{\small
\begin{figure}[h]
\begin{center}
\begin{tabular}{c|c|c|c|}\hline
$R(D_1)$&$A$ & $B$ & $C$\\ \hline
$t_1$&$a_1$ & ${\bf b_{12}}$ & $c_{1}$ \\
$t_2$&$a_2$ & ${\bf b_{12}}$ & $c_{2}$\\
$t_3$&$a_3$ & $b_{3}$ & $c_3$\\ \cline{2-4}
\end{tabular}\hspace{4mm}
\begin{tabular}{c|c|c|c|}\hline
$R(D_2)$&$A$ & $B$ & $C$\\ \hline
$t_1$&$a_1$ & $b_{12}$ & ${\bf c_{12}}$ \\
$t_2$&$a_2$ & $b_{12}$ & ${\bf c_{12}}$\\
$t_3$&$a_3$ & $b_{3}$ & $c_3$\\ \cline{2-4}
\end{tabular}
\end{center} \vspace{-3mm}
\caption{Instances $D_1$ and $D_2$, resp.} \label{fig:d1}\vspace{-4mm}
\end{figure} }


Now, enforcing $\varphi_1$ on $t_1,t_3$ in $D_2$ makes the tuples get the same value for attribute $R[B]$, as shown in instance $D_3$ (cf.~Figure~\ref{fig:d3}). At this stage we have broken the  equality of $t_1[B],t_2[B]$ we had  in $D_2$, as shown underlined in Figure~\ref{fig:d3}. This is  a crucial point in relation to the SFAI property: \ $\varphi_1$ is still applicable on $D_3$ with $t_1, t_2$,  because there are no MDs with attribute $R[A]$ in their
RHSs that could destroy the initial similarities that held in $D_0$, in particular $t_1[A]=a_1 \approx a_3=t_3[A]$: they keep  holding along the enforcement path. \ignore{\red{This is because there is no blocking-MDs $m$ with attribute $\nit{Name}$,  or $\nit{Aff}$ in $\nit{RHS}(m)$  ($\Sigma$ is SFAI on $D_0$)}.} So, enforcing $\varphi_1$ makes $t_1[B],t_2[B]$ identical again, as shown in instance $D_4$ (cf.~Figure~\ref{fig:d3}).

Notice that the initial similarities of attribute values we have in the initial instances are not destroyed
later along a chase sequences.  This is a general property for SFAI combinations.

{\small

\begin{figure}[h]
\begin{center}
\begin{tabular}{c|c|c|c|}\hline
$R(D_3)$&$A$ & $B$ & $C$\\ \hline
$t_1$&$a_1$ & $\underline{{\bf b_{123}}}$ & $c_{12}$ \\
$t_2$&$a_2$ & $\underline{b_{12}}$ & $c_{12}$\\
$t_3$&$a_3$ & ${\bf b_{123}}$ & $c_3$\\ \cline{2-4}
\end{tabular}\hspace{4mm}
\begin{tabular}{c|c|c|c|}\hline
$R(D_4)$&$A$ & $B$ & $C$\\ \hline
$t_1$&$a_1$ & ${\bf b_{123}}$ & $c_{12}$ \\
$t_2$&$a_2$ & ${\bf b_{123}}$ & $c_{12}$\\
$t_3$&$a_3$ & $b_{123}$ & $c_3$\\ \cline{2-4}
\end{tabular}
\end{center}\vspace{-3mm}
\caption{Instances $D_3$ and $D_4$, resp.}\label{fig:d3} \vspace{-4mm}
\end{figure} }

\vspace*{0.3cm}

Next, applying $\varphi_2$ on  $t_2, t_3$ in  $D_4$ makes the tuples get the same value for attribute $R[C]$, as shown in instance $D_5$ (cf.~Figure~\ref{fig:d4}). Enforcing  $\varphi_2$ on $t_1$, $t_2$ in $D_5$ results in instance $D_6$, as shown in Figure~\ref{fig:d4}. No further applications of MDs are possible, and we have reached a stable instance.

\begin{figure}[h]
\begin{center}
\begin{tabular}{c|c|c|c|}\hline
$R(D_5)$&$A$ & $B$ & $C$\\ \hline
$t_1$&$a_1$ & $ b_{123}$ & $ c_{12}$ \\
$t_2$&$a_2$ & $b_{123}$ & ${\bf c_{123}}$\\
$t_3$&$a_3$ & $b_{123}$ & ${\bf c_{123}}$\\ \cline{2-4}
\end{tabular}\hspace{4mm}
\begin{tabular}{c|c|c|c|}\hline
$R(D_6)$&$A$ & $B$ & $C$\\ \hline
$t_1$&$a_1$ & $b_{123}$ & ${\bf c_{123}}$ \\
$t_2$&$a_2$ & $b_{123}$ & ${\bf c_{123}}$\\
$t_3$&$a_3$ & $b_{123}$ & $c_{123}$\\ \cline{2-4}
\end{tabular}
\end{center}\vspace{-3mm}
\caption{Instances $D_5$ and $D_6$, resp.}\label{fig:d4} \vspace{-4mm}
\end{figure}

\vspace*{0.3cm}

Actually, $D_6$ is the only instance that can be reached through any chase sequence. For example,  we will now show another chase sequence leading to the same clean instance $D_6$.

The above chase sequence started applying $\varphi_1$ with $t_1, t_2$. We could have started with enforcing $\varphi_1$ on $t_1,t_3$ in $D_0$. This makes  the tuples get the same value for attribute $R[B]$, as shown in instance $D''_1$ (cf.~Figure~\ref{fig:d312}). Next, enforcing  $\varphi_1$ on $t_1$, $t_2$ in $D''_1$ results in instance $D''_2$,  where $t_1,t_2$ have identical values for attribute $R[B]$, as shown in Figure~\ref{fig:d312}.

\vspace*{0.3cm}
{\small
\begin{figure}[h]
\begin{center}
\begin{tabular}{c|c|c|c|}\hline
$R(D''_1)$&$A$ & $B$ & $C$\\ \hline
$t_1$&$a_1$ & ${\bf b_{13}}$ & $c_{1}$ \\
$t_2$&$a_2$ & $b_{2}$ & $c_{2}$\\
$t_3$&$a_3$ & ${\bf b_{13}}$ & $c_3$\\ \cline{2-4}
\end{tabular}\hspace{4mm}
\begin{tabular}{c|c|c|c|}\hline
$R(D''_2)$&$A$ & $B$ & $C$\\ \hline
$t_1$&$a_1$ & $\underline{{\bf b_{123}}}$ & $c_{1}$ \\
$t_2$&$a_2$ & ${\bf b_{123}}$ & $c_{2}$\\
$t_3$&$a_3$ & $\underline{b_{13}}$ & $c_3$\\ \cline{2-4}
\end{tabular}
\end{center}\vspace{-3mm}
\caption{Instances $D''_1$ and $D''_2$, resp.}\label{fig:d312} \vspace{-4mm}
\end{figure} }

\vspace*{0.3cm}

Again, we have broken the  equality of $t_1[B],t_3[B]$ we had  in $D''_1$, as shown underlined in Figure~\ref{fig:d312}. MD $\varphi_1$ is still applicable on $D''_2$ with $t_1, t_3$. So, enforcing $\varphi_1$ makes $t_1[B],t_3[B]$ identical again, as shown in instance $D''_3$ (cf.~Figure~\ref{fig:d3121}).

\vspace*{0.3cm}

{\small
\begin{figure}[h]
\begin{center}
\begin{tabular}{c|c|c|c|}\hline
$R(D''_3)$&$A$ & $B$ & $C$\\ \hline
$t_1$&$a_1$ & ${\bf b_{123}}$ & $c_{1}$ \\
$t_2$&$a_2$ & $b_{123}$ & $c_{2}$\\
$t_3$&$a_3$ & ${\bf b_{123}}$ & $c_3$\\ \cline{2-4}
\end{tabular}\hspace{4mm}
\begin{tabular}{c|c|c|c|}\hline
$R(D''_4)$&$A$ & $B$ & $C$\\ \hline
$t_1$&$a_1$ & $ b_{123}$ & ${\bf c_{1}}$ \\
$t_2$&$a_2$ & $b_{123}$ & $ {\bf c_{23}}$\\
$t_3$&$a_3$ & $ b_{123}$ & ${\bf c_{23}}$\\ \cline{2-4}
\end{tabular}
\end{center}\vspace{-3mm}
\caption{Instances $D''_3$ and $D''_4$, resp.}\label{fig:d3121} \vspace{-4mm}
\end{figure} }

\vspace*{0.3cm}

Next, applying $\varphi_2$ on  $t_2, t_3$ in  $D''_3$, makes the  tuples get the same value for attribute $R[C]$, as shown in instance $D''_4$ (cf.~Figure~\ref{fig:d3121}). Now, enforcing $\varphi_2$ on $t_1,t_3$ in $D''_4$ makes the  tuples get the same value for attribute $R[C]$, as shown in instance $D''_5$ (cf.~Figure~\ref{fig:d31211}). Enforcing  $\varphi_2$ on $t_1$, $t_2$ in $D''_5$ results in instance $D_6$ which we had obtained before through a different chase sequence.

{\small
\begin{figure}[h]
\begin{center}
\begin{tabular}{c|c|c|c|}\hline
$R(D''_5)$&$A$ & $B$ & $C$\\ \hline
$t_1$&$a_1$ & $ b_{123}$ & ${\bf c_{123}}$ \\
$t_2$&$a_2$ & $b_{123}$ & $ c_{23}$\\
$t_3$&$a_3$ & $b_{123}$ & ${\bf c_{123}}$\\ \cline{2-4}
\end{tabular}
\end{center}\vspace{-3mm}
\caption{Instance $D''_5$}\label{fig:d31211} \vspace{-4mm}
\end{figure} }

\vspace*{0.3cm}

Actually, no matter in what order the MDs are enforced in this case, the final, clean instance will be $D_6$, which is due to $(\Sigma, D_0)$ having the SFAI property.
\boxtheorem
\end{example}

We illustrated the definition of SFAI with a classical set of MDs.

\begin{example}\label{ex:relationalMD21}  Consider the set of  relational MDs $\Sigma = \{\varphi_1,\varphi_2\}$ with:
{\small
\begin{eqnarray*}
\varphi_1\!:&& \ \nit{Author}(t_1,x_1,y_1, p_1,\nit{bl}_1) \ \wedge \  \nit{Author}(t_2,x_2,y_2, p_2,\nit{bl}_2)  \ \wedge \  x_1 \approx x_2  \ \wedge \nonumber \\ && \hspace{1cm} \nit{Paper}(t_3,p_1,z_1, w_1, \nit{bl}_4) \ \wedge \
 \nit{Paper}(t_4,p_2,z_2, w_2, \nit{bl}_4)  \  \longrightarrow \  \nit{bl}_1 \doteq \nit{bl}_2.\hspace{1cm} \label{eq:blmd2}\\
 \varphi_2\!:&& \nit{Paper}(t_1,p_1,z_1, w_1, \nit{bl}_1) \ \wedge \   \nit{Paper}(t_2, p_2,z_2, w_2, \nit{bl}_2) \ \wedge \  z_1 \approx z_2 \ \wedge \\
 &&\hspace{1cm}\nit{Author}(t_3,x_1,y_1,p_1,\nit{bl}_3) \ \wedge \  \nit{Author}(t_4, x_2,y_2,p_2,\nit{bl}_3)    \  \longrightarrow \  \nit{bl}_1 \doteq \nit{bl}_2.\nonumber
 \end{eqnarray*} }
Assume  that the only similarities that hold in the data domain $U$, apart from equalities, are: \ $n_2 \approx n_3$ and $\nit{title}_1 \approx \nit{title}_3$.
Consider the  initial instance $D_0$:\\

{\footnotesize
\hspace*{-5mm}\begin{tabular}{c|c|c|c|c|}\hline
$\nit{Author}(D_0)$&$\nit{Name}$ &  $\nit{Aff}$&$\nit{PID}$& $Bl \#$\\ \hline
$t_1$& $n_1$ & $a_1$ &$120$ &$250$\\
$t_2$& $n_2$ & $a_2$ &$121$ &$251$\\
$t_3$& $n_3$ & $a_3$ &$122$ &$252$\\\cline{2-5}
\end{tabular}~~~
\begin{tabular}{c|c|c|c|c|}\hline
$\nit{Paper}(D_0)$&$\nit{PID}$ & $\nit{Title}$ &$\nit{Key}$ & $Bl\#$\\ \hline
$t_4$&$120$ & $\nit{title}_1$ & $k_1$ & $302$\\
$t_5$&$122$ & $\nit{title}_2$ & $k_2$ & $300$\\
$t_6$&$121$ & $\nit{title}_3$ & $k_3$ & $300$\\\cline{2-5}
\end{tabular}
}

\vspace{3mm}

Here, to check the SFAI property, we find {\em two cases of interaction}:

(1) $\nit{ALHS}(\varphi_2) \cap \nit{ARHS}(\varphi_1) = \{ \nit{Author}[\nit{Bl \#}]\}$,

(2) $\nit{ALHS}(\varphi_1) \cap \nit{ARHS}(\varphi_2) = \{ \nit{Paper}[\nit{Bl \#}]\}$.

We have to check both cases   according to condition (b) in Definition \ref{def:SFAInew}. For example, for the first case, $(\Sigma, D_0)$ is not SFAI if there are $S_1,S_2 \subseteq D_0$ with a tuple $\nit{Author}(\bar{a}) \in  S_1 \cap S_2$, such that $\nit{LHS}(\varphi_1)$ is true in $S_1$ and  $\nit{LHS}(\varphi_2)$ is true in $S_2$.
\boxtheorem
\end{example}

In the general relational case of MDs, one would wonder how difficult is  checking the SFAI property. First, notice that
 only the active domain, $\nit{Adom}(D)$, of the initial instance $D$ matters  for  condition (b),  because  $S_1,S_2$ are subsets of $D$.
Actually, checking the SFAI property is  decidable,  because, for condition (a), a finite set of MDs has to be checked, for interaction; and, for condition (b),  $\nit{Adom}(D)$  is finite. Even more, the test can be performed in polynomial
time in the size of $D$ (i.e.~{\em in data}), by posing one Boolean conjunctive query (BCQ) (with similarity built-ins) for each case of interaction between any two MDs in $\Sigma$.\footnote{The MD $R\left[B\right] \approx R\left[B\right] \rightarrow R\left[B\right] \doteq R\left[B\right]$ interacts with itself, and gives rise to one SFAI test (one query). The interacting
MDs \ $R\left[B\right] \approx R\left[B\right] \rightarrow R\left[A\right] \doteq R\left[A\right], \ R\left[A\right] \approx R\left[A\right] \rightarrow R\left[B\right] \doteq R\left[B\right]$ give rise to two SFAI tests (two queries).}

 If one of those queries
gets the value $\nit{true}$ in $D$, the SFAI property does not hold. We illustrate this claim with an example.

\begin{example}\label{ex:relationalMD2}    (ex. \ref{ex:relationalMD21} cont.) For the first case of interaction between the MDs, the following BCQ is posed to $D_0$:
\begin{eqnarray*}\mc{Q}_{\varphi_1,\varphi_2}\!:&&    \exists \bar{t} \ \exists \bar{x} \ \exists \bar{y} \ \exists \bar{p} \ \exists \bar{\nit{bl}} \ \exists \bar{z} \ \exists \bar{w} \  (\nit{Author}(t_1,x_1,Y_1, p_3,\nit{bl}_1) \ \wedge \  \\
 &&~~~~~\nit{Author}(t_2,x_2, y_2, p_4, \nit{bl}_2)  \ \wedge \  \nit{Paper}(t_3,p_3,z_3, w_3, \nit{bl}_3)\ \wedge \  \\
 &&~~~~~\nit{Paper}(t_4,p_4,z_4, w_4, \nit{bl}_3) \ \wedge \  \nit{Author}(t_5,x_5,y_5, p_6,\nit{bl}_2) \ \wedge \   \\
 &&~~~~~\nit{Paper}(t_6,p_6,z_6, w_6, \nit{bl}_4) \ \wedge \  x_1 \approx x_2 \ \wedge \  z_4 \approx z_6).
\end{eqnarray*}
$\mc{Q}_{\varphi_1,\varphi_2}$ takes the value $\nit{false}$ in $D_0$, then this case (case (1) in Example \ref{ex:relationalMD21}) does not lead to a violation of the SFAI property.

For the second case of interaction, we consider the following BCQ:
\begin{eqnarray*}\mc{Q}_{\varphi_2,\varphi_1}\!:&&  \exists \bar{t} \ \exists \bar{x} \ \exists \bar{y} \ \exists \bar{p} \ \exists \bar{\nit{bl}} \ \exists \bar{z} \ \exists \bar{w} \  (\nit{Author}(t_1,x_1,y_1, p_3,\nit{bl}_1) \ \wedge \ \\
&&~~~~~\nit{Author}(t_2,x_2, y_2, p_4, \nit{bl}_1) Â \ \wedge \ Â \nit{Paper}(t_3,p_3,z_3, w_3, \nit{bl}_3)\ \wedge \ \\
&&~~~~~\nit{Paper}(t_4,p_4,z_4, w_4, \nit{bl}_4) \ \wedge \ Â \nit{Author}(t_5,x_5,y_5, p_6,\nit{bl}_2) \ \wedge \  \\
&&~~~~~\nit{Paper}(t_6,p_6,z_6, w_6, \nit{bl}_4) \ \wedge \ \nit{Author}(t_7,x_7,y_7, p_6,\nit{bl}_7) \ \wedge \ \\
&&~~~~~ x_2 \approx x_7 \ \wedge \ z_4 \approx z_3).
\end{eqnarray*}
$\mc{Q}_{\varphi_2,\varphi_1}$ also takes the value $\nit{false}$ in $D_0$. Then, \ $(\Sigma, D_0)$ is SFAI.


For a negative example of SFAI, with the same MDs $\varphi_1, \varphi_2$, consider a different initial instance $D_1$, :

\vspace*{0.4cm}
{\footnotesize
\hspace*{-5mm}\begin{tabular}{c|c|c|c|c|}\hline
$\nit{Author}(D_1)$&$\nit{Name}$ & Â $\nit{Aff}$&$\nit{PID}$& $Bl \#$\\ \hline
$t_1$& $n_1$ & $a_1$ &$120$ &$250$\\
$t_2$& $n_2$ & $a_2$ &$121$ &$250$\\
$t_3$& $n_3$ & $a_3$ &$122$ &$252$\\
$t_4$& $n_4$ & $a_4$ &$121$ &$253$\\\cline{2-5}
\end{tabular}~~~
\begin{tabular}{c|c|c|c|c|}\hline
$\nit{Paper}(D_1)$&$\nit{PID}$ & $\nit{Title}$ &$\nit{Key}$ & $Bl\#$\\ \hline
$t_5$&$120$ & $\nit{title}_1$ & $k_1$ & $302$\\
$t_6$&$122$ & $\nit{title}_2$ & $k_2$ & $300$\\
$t_7$&$121$ & $\nit{title}_3$ & $k_3$ & $300$\\\cline{2-5}
\end{tabular}
}

\vspace*{0.4cm}

In this case, $(\Sigma, D_1)$ is not SFAI, because the query $\mc{Q}_{\varphi_2,\varphi_1}$ above takes the value $\nit{true}$ in $D_1$.
\boxtheorem
\end{example}
As shown in Example \ref{ex:newclass1} with a set of classical  MDs $\Sigma$, when $(\Sigma, D)$ is SFAI, the initial similarities that held in $D$ can not be destroyed during a complete chase sequence. In particular,  the initial similarities keep  holding along the enforcement path.  The same holds for relational SFAI combinations. As a consequence, enforcements of SFAI combinations behave similarly to the case of non-interacting MDs.
Actually, it is possible to prove that SFAI combinations have the SCI property. Even more, it is possible to automatically rewrite  {\em answer set programs} \cite{brewka} that specify the clean instances obtained with general sets of MDs
\cite{Bahmani12} into Datalog programs with stratified negation \cite{corrFlairs17}, which have a single ({\em standard}) model that coincides with the single clean instance.

\end{document}